\documentclass[aps,twocolumn,nopacs,superscriptaddress,preprintnumbers,prb]{revtex4-1}
\usepackage{graphicx}
\usepackage{bm}
\usepackage{bbold}
\usepackage{amsmath}

\usepackage{epsfig}
\usepackage{physics}
\usepackage{epstopdf}
\usepackage[colorlinks=true,
            linkcolor = blue,
            urlcolor  = blue,
            citecolor = blue,
            anchorcolor = blue]{hyperref}
\usepackage{url}
\usepackage{color}
\usepackage{multirow}
\usepackage{comment}
\usepackage{mathrsfs}
\usepackage{subfigure}
\begin{document}

\title{Leading components and pressure-induced color changes in N-doped lutetium hydride}

\author{Xiangru Tao$^{\star}$}
\affiliation{MOE Key Laboratory for Non-equilibrium Synthesis and Modulation of Condensed Matter, Shaanxi Province Key Laboratory of Advanced Functional Materials and Mesoscopic Physics, School of Physics, Xi'an Jiaotong University, 710049, Xi'an, Shaanxi, P.R.China}

\author{Aiqin Yang$^{\star}$}
\affiliation{MOE Key Laboratory for Non-equilibrium Synthesis and Modulation of Condensed Matter, Shaanxi Province Key Laboratory of Advanced Functional Materials and Mesoscopic Physics, School of Physics, Xi'an Jiaotong University, 710049, Xi'an, Shaanxi, P.R.China}

\author{Shuxiang Yang}
\affiliation{Zhejiang laboratory, Hangzhou, Zhejiang, P.R.China}

\author{Yundi Quan}
\email{yundi.quan@gmail.com}
\affiliation{MOE Key Laboratory for Non-equilibrium Synthesis and Modulation of Condensed Matter, Shaanxi Province Key Laboratory of Advanced Functional Materials and Mesoscopic Physics, School of Physics, Xi'an Jiaotong University, 710049, Xi'an, Shaanxi, P.R.China}

\author{Peng Zhang}
\email{zpantz@mail.xjtu.edu.cn\\}
\affiliation{MOE Key Laboratory for Non-equilibrium Synthesis and Modulation of Condensed Matter, Shaanxi Province Key Laboratory of Advanced Functional Materials and Mesoscopic Physics, School of Physics, Xi'an Jiaotong University, 710049, Xi'an, Shaanxi, P.R.China}

\begin{abstract}
Recent experimental study by Dias {\it et al.} claims to have discovered room-temperature superconductivity in lutetium-nitrogen-hydrogen system at 1 GPa [Nature 615, 244 (2023)], which sheds light on the long-held dream of ambient superconductivity. However, all follow-up experiments found no evidence of superconductivity. The compositions and the crystal structures of the lutetium-nitrogen-hydrogen system remain unknown. By employing the density functional theory based structure prediction algorithm, we suggest that in lutetium-nitrogen-hydrogen the major component is LuH$_2$ (Fm$\Bar{3}$m), together with minor LuN (Fm$\Bar{3}$m). The blue LuH$_2$ at ambient pressure will turn into purple and red color at higher pressures, possibly accompanied by the formation of vacancies at hydrogen-sites. In LuH$_2$ and LuN, the density of states at the Fermi level is dominated by the Lu-5d orbitals, while those from hydrogen and nitrogen are very small, leading to the absence of superconductivity in these two compounds. Nitrogen-doping to LuH$_2$ fails to enhance the superconductivity as well. In this work, we identify the leading components in N-doped lutetium hydride, explain its intriguing color changes under pressure, and elucidate why superconductivity is absent in the follow-up experiments.
\end{abstract}


\maketitle
\section{Introduction}
In 2004, Ashcroft proposed the idea of pre-compressing the hydrogen atoms by their surrounding atoms in hydrides, so that the required physical pressure for superconductivity can be reduced by the compensated chemical pressure \cite{Ashcroft.PRL2004}. Thanks to the efficient cooperation among the density functional theory (DFT) based crystal structure prediction methods \cite{AIRSS,CALYPSO,USPEX}, the Eliashberg theory for calculations of the electron-phonon type superconductors \cite{Eliashberg} and the diamond anvil dell (DAC) experiments \cite{DAC}, scientific community has witnessed the booming discoveries of high temperature superconducting hydrides in the last decade \cite{2021Roadmap, Pickett.RevModPhys2023}. In 2014, Duan $\textit{et al.}$ \cite{Duan.SciRept2014} predicted that H$_3$S will superconduct with $T_{\text{c}}$ $\approx$ 204 K at 200 GPa , which was later confirmed by the DAC experiment of Drozdov $\textit{et al.}$ \cite{Drozdov.Nature2015}. A series of rare earth hydrides were also predicted to be superconductors \cite{Peng.PRL2017}, including LaH$_{10}$ with $T_{\text{c}}$ $\approx$ 274 - 286 K at 210 GPa \cite{Liu.PNAS2017}, which was confirmed by later experiments as well \cite{Geballe.Angew2018, Somayazulu.PRL2019, Drozdov.Nature2019}. Up to date, the discovered hydrogen-rich superconductors range from binary hydrides like sulfur hydrides \cite{Duan.SciRept2014,Drozdov.Nature2015}, rare earth hydrides \cite{Peng.PRL2017, Drozdov.Nature2019,Errea2020}, alkali metal hydrides \cite{Zurek.JCP2019}, alkali earth metal hydrides \cite{Zurek.CIC2017}, transition metal hydrides \cite{Gao.MTP2021} to ternary hydrides \cite{2021Roadmap}. However, although the superconducting transition temperature of these hydrides could be pretty high or even close to room temperature, they are usually not stable unless in a very high pressure environment, typically at a few hundreds of GPa. This greatly limits the potential application of these high-temperature superconductors. 

Recently, Dasenbrock-Gammon $\textit{et al.}$ \cite{Dias.Nature2023} claims that they found a maximum $T_{\text{c}}$ of 294 K in the lutetium-nitrogen-hydrogen (Lu-N-H) system at 1 GPa, which is more than two orders of magnitude lower than the pressures of maximum $T_{\text{c}}$ in typical high temperature hydrides, such as H$_3$S and LaH$_{10}$. Indeed the exploration of ambient superconductivity in rare earth hydrides is promising, especially considering that LuH$_6$ (Im$\bar{3}$m) was predicted to be superconducting with $T_{\text{c}}$ $\approx$ 273 K at 100 GPa \cite{Duan.CPL2021}, and the recent experimentally discovered superconductivity of $T_{\text{c}}$ $\approx$ 71 K at 218 GPa in lutetium polyhydride \cite{Jin.SciChina2023}. The breakthrough of near ambient superconductivity in Lu-N-H inspires immediate interest in examining whether the discovery could be reproduced \cite{Jin.Nature2023, Shan.CPL2023, Zhao.2023, Wang.arxiv2023, Ming.arxiv2023, Zhang.arxiv2023, Xing.arxiv2023, Cao1.arxiv2023, Cai.arxiv2023, Moulding.arxiv2023, Cao2.arxiv2023, ZLiu.arxiv2023, Liu.arxiv2023, Huo.arxiv2023, Xie.CPL2023, Hilleke.arxiv2023, Ferreira.arxiv2023, Sun.arxiv2023, Lucrezi.arxiv2023, Lu.arxiv2023, Kim.arxiv2023, Dangic.arxiv2023}. 
Unfortunately, all follow-up experiments cannot find any sign of superconductivity. Shan $\textit{et al.}$ \cite{Shan.CPL2023} found no evidence of superconductivity in LuH$_2$ under pressures up to 7.7 GPa and temperatures down to 1.5 K . Ming $\textit{et al.}$ \cite{Ming.arxiv2023}, Zhang $\textit{et al.}$ \cite{Zhang.arxiv2023} and Xing $\textit{et al.}$ \cite{Xing.arxiv2023} claim the absence of superconductivity in nitrogen-doped lutetium hydride LuH$_{2\pm x}$N$_y$ under pressures up to 50.5 GPa and temperatures down to 1.8 K. Cai $\textit{et al.}$ \cite{Cai.arxiv2023} also claim no evidence of superconductivity in Lu-N-H sample, which is prepared using the same strategy as Dias $\textit{et al.}$ Several theoretical calculations have been done for the Lu-N-H system as we conduct this work, but no evidence for the existence of room temperature superconductivity was found \cite{Liu.arxiv2023, Huo.arxiv2023, Xie.CPL2023, Hilleke.arxiv2023, Ferreira.arxiv2023, Sun.arxiv2023, Lucrezi.arxiv2023, Lu.arxiv2023, Kim.arxiv2023}. 
Thermodynamically stable Lu-N-H ternary compounds have not been discovered in most structure searches \cite{Liu.arxiv2023, Huo.arxiv2023, Xie.CPL2023, Hilleke.arxiv2023}, except Lu$_{4}$N$_{2}$H$_{5}$ (P2/c) being suggested by Ferreira $\textit{et al.}$ \cite{Ferreira.arxiv2023}. Superconductivity in Lu-N-H was believed to be possibly related to  LuH$_3$ (Fm$\Bar{3}$m) \cite{Dias.Nature2023}, which is dynamical unstable at low pressure. Sun $\textit{et al.}$ \cite{Sun.arxiv2023}'s calculations suggest that Lu$_8$H$_{21}$N (P$\Bar{4}$m2), derived from nitrogen-doped LuH$_3$ (Fm$\Bar{3}$m) with H-vacancies, could be dynamically stable at 1 GPa.
Huo $\textit{et al.}$ \cite{Huo.arxiv2023} found that including of nitrogen-atoms into LuH$_3$ (Fm$\Bar{3}$m) would decrease its dynamical stability, where the minimum dynamically stable pressure increased from 25 GPa to 70 GPa as the doping concentration increased from 0\% to 2\%. Huo $\textit{et al.}$ \cite{Huo.arxiv2023}'s electron-phonon calculations suggest superconductivity in nitrogen-doped LuH$_3$, with maximum $T_{\text{c}}$ = 22 K at 30 GPa, still far below the ambient temperature. Lucrezi $\textit{et al.}$ \cite{Lucrezi.arxiv2023} suggest that temperature and quantum anharmonic phonons could help to stabilize LuH$_3$ (Fm$\Bar{3}$m), with the predicted superconducting critical temperature $T_{\text{c}}$ in the range of 50 - 60 K. 

Although these subsequent experiments and calculations \cite{Shan.CPL2023, Zhao.2023, Wang.arxiv2023, Ming.arxiv2023, Zhang.arxiv2023, Xing.arxiv2023, Cao1.arxiv2023, Cai.arxiv2023, Moulding.arxiv2023, Cao2.arxiv2023, Liu.arxiv2023, Huo.arxiv2023, Xie.CPL2023, Sun.arxiv2023, Hilleke.arxiv2023, Lucrezi.arxiv2023, Lu.arxiv2023, Kim.arxiv2023} have resolved several important issues concerning the Lu-N-H, a few critical questions remain open. First, is LuH$_2$ the only component of the Lu-N-H sample? Second, if not, what are the other components, or at least the leading components of Lu-N-H?
Although Dasenbrock-Gammon $\textit{et al.}$ \cite{Dias.Nature2023} indicates there are two distinct hydrides with the same face-centered cubic (FCC) sublattice of Lu, the exact compositions of the Lu-N-H sample are still unknown. 
Third, Dasenbrock-Gammon $\textit{et al.}$ \cite{Dias.Nature2023} found blue to red color change in Lu-N-H, which was observed as well by others in the compressed LuH$_2$ \cite{Shan.CPL2023, Zhao.2023, Ming.arxiv2023} and LuH$_{2\pm x}$N$_y$ \cite{Zhang.arxiv2023, Xing.arxiv2023}, although the transition pressures are diverse. What is the origin of the blue to red color change?  

In this work, we report thorough searches of Lu-H and Lu-N binary systems employing the DFT based structure prediction package USPEX \cite{USPEX}. By comparing the derived X-ray diffraction (XRD) pattern with the experimental results of Dasenbrock-Gammon $\textit{et al.}$ \cite{Dias.Nature2023}, together with the thermodynamical stability analysis and the phonon spectra calculations, we suggest that in Lu-N-H the dominant component is LuH$_2$ (Fm$\Bar{3}$m), together with minor LuN (Fm$\Bar{3}$m). Our optical reflectivity calculations indicate that the blue to purple then to red color changes in Lu-N-H originate from the increased pressure on LuH$_2$ (Fm$\Bar{3}$m), probably being assisted by the vacancy formation at the hydrogen-sites. Neither the stoichiometric LuH$_2$ (Fm$\Bar{3}$m), the stoichiometric LuN (Fm$\Bar{3}$m), nor the nitrogen-doped LuH$_2$ (Fm$\Bar{3}$m) are superconducting since the density of states (DOS) of hydrogen at the Fermi level in these compounds are extremely small.

\section{Methods}
Our USPEX variable-composition search of the Lu-H and the Lu-N binary systems starts with 90 randomly initiated structures comprising of up to 40 atoms in the primitive cell. 50 structures are generated in each generation. The projector augmented-wave \cite{PAW,Kresse.PRB1999} DFT based Vienna ab initio simulation package (VASP) \cite{Kresse.PRB1993,Kresse.PRB1994,Kresse.PRB1996} package is employed with the convergence criteria set as 0.03 eV/Å and 10$^{-6}$ eV for structure relaxations and total energy calculations. The generalized gradient approximation of Perdew-Burke-Ernzerhof \cite{PBE} exchange-correlation functional is used. The plane-wave energy cutoff of 400 eV, the Methfessel-Paxton electronic smearing \cite{Methfessel.PRB1989}, and the $\Gamma$-centered k-point meshes with a resolution of 2$\pi$ × 0.1 Å$^{-1}$ are used to ensure the convergence of total energy. In the fixed-composition search, each generation has 30 structures, except that 50 structures are randomly generated at the first generation. All structures in the fixed-composition search are relaxed with higher precision using the energy cutoff of 500 eV and the k-point mesh of 2$\pi$ × 0.03 Å$^{-1}$.

The electronic structures and the phonon calculations are performed using the pseudopotential code QUANTUM-ESPRESSO (QE) package \cite{QE}. The VASP-optimized structures are optimized again with QE. We employ the optimized norm-conversing pseudopotential proposed by Hamann with the valence electron configurations of Lu-5$s^2$5$p^6$4$f^{14}$5$d^1$6$s^2$, H-1$s^1$ and N-2$s^2$2$p^3$\cite{ONCV}. The plane-wave kinetic-energy cutoff and the charge density energy cutoff are 100 Ry and 400 Ry, respectively. The Methfessel-Paxton smearing width of 0.02 Ry is used. 24 × 24 × 24 k-point grids in Brillouin zone are adopted for the self-consistent electron-density calculations. The dynamic matrix and the electron-phonon coupling (EPC) constant $\lambda$ are calculated using the density-functional perturbation theory \cite{DFPT} with an 8 × 8 × 8 mesh of q-points. 
The superconducting transition temperature is estimated following the Allen-Dynes modified McMillan equation \cite{Allen.PR1975},
\begin{eqnarray}
T_{\text{c}} = \frac{\omega_{\text{log}}}{1.2} \exp \left ( -\frac{1.04(1+\lambda)}{\lambda - \mu^\ast\left (1+0.62 \lambda 
 \right ) }\right ),
\end{eqnarray}
in which $\lambda$ is the average EPC parameter, $\omega_{\text{log}}$ is the logarithmic average frequency, and the Coulomb pseudopotential \cite{Morel.PR1962} $\mu^{*}$ = 0.10.

Employing the all-electron WIEN2K package \cite{WIEN2K}, the optical reflectivity is calculated. The cutoff energy for core states is $-$6.0 Ry and $R_{\text{mt}}\cdot K_{\text{max}}=7$, where $R_{\text{mt}}$ is the smallest atomic sphere radius in the unit cell and $K_{\text{max}}$ is the magnitude of the largest $K$ vector. The convergence criteria of charge and energy are set as $10^{-5}$ $e$ and $10^{-5}$ Ry, respectively. Up to 10000 k-points are used in calculating the optical reflectivity. The plasma frequencies are listed in the supplementary information.

\section{Results}
\subsection{Convex hull and XRD}
\begin{figure*}[t]
    \centering
    \includegraphics[width=1.0\textwidth]{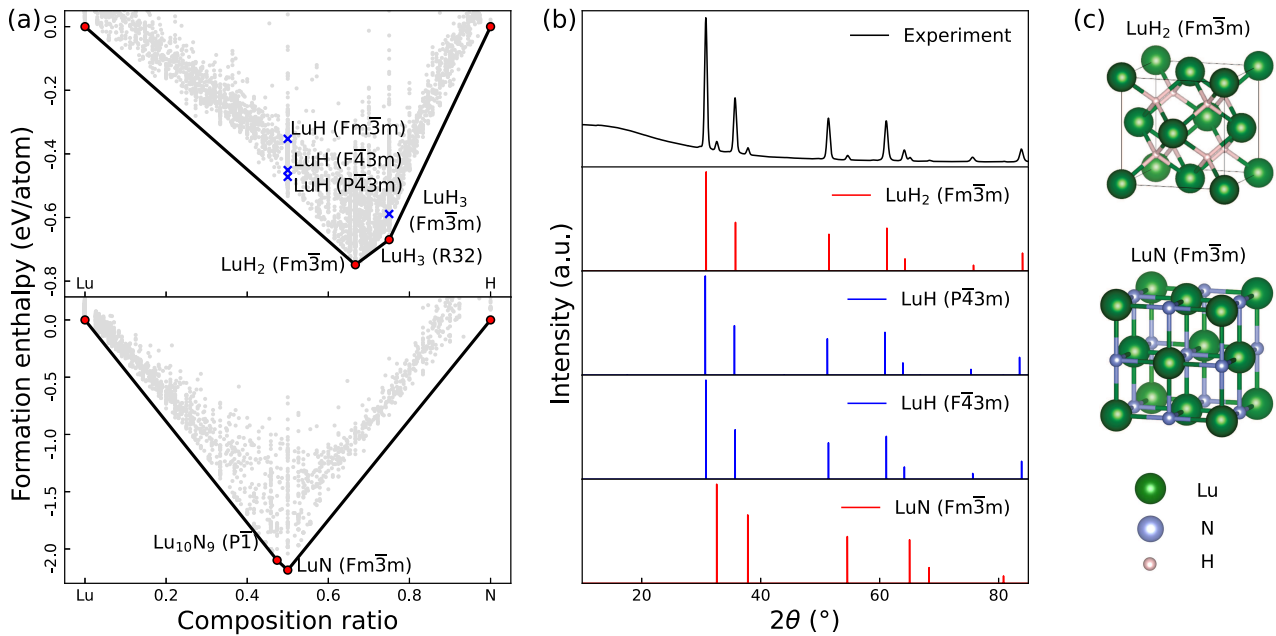}
    \caption{(Color online) (a) Formation enthalpy of the predicted Lu–H and Lu–N structures at 2 GPa. Red-filled circles on the convex hull, LuH$_2$ (Fm$\Bar{3}$m), LuH$_3$ (R32), Lu$_{10}$N$_9$ (P$\Bar{1}$) and LuN (Fm$\Bar{3}$m), represent the thermodynamically stable structures. Blue crosses, LuH (P$\Bar{4}$3m), LuH (F$\Bar{4}$3m), LuH (Fm$\Bar{3}$m) and LuH$_3$(Fm$\Bar{3}$m), represent the thermodynamically meta-stable structures with FCC Lu-sublattice. Other meta-stable structures are marked by grey dots. The composition ratio is defined as $N_{\text{X}}/(N_{\text{Lu}}+N_{\text{X}}$), where $N_{\text{X}}$ and $N_{\text{Lu}}$ represent the number of atoms in the formula unit and X= H, N. (b) The XRD patterns of four candidate structures, LuH$_2$ (Fm$\Bar{3}$m), LuH (P$\Bar{4}$3m), LuH (F$\Bar{4}$3m), and LuN (Fm$\Bar{3}$m), in comparison with the experimental results of Dasenbrock-Gammon $\textit{et al.}$ \cite{Dias.Nature2023} at ambient pressure. (c) Crystal structures of LuH$_2$ (Fm$\Bar{3}$m) and LuN (Fm$\Bar{3}$m), as the leading components of Lu-N-H.}
    \label{hull}
\end{figure*}

The structure prediction results of the Lu-H and the Lu-N binary systems are summarized in Fig.~\ref{hull}a. In the Lu-H binary system at 2 GPa, we found two thermodynamically stable stoichiometric Lu-H binary structures, LuH$_2$ (Fm$\Bar{3}$m) and LuH$_3$ (R32), at the convex hull. 
The two structures are also discovered by Liu $\textit{et al.}$ \cite{Liu.arxiv2023}, but thermodynamically stable LuH$_3$ (P$\Bar{3}$c1) and LuH$_3$ (P6$_3$) are suggested in calculations by Xie $\textit{et al.}$ \cite{Xie.CPL2023} and Ferreira $\textit{et al.}$ \cite{Ferreira.arxiv2023}, respectively. 
Dasenbrock-Gammon $\textit{et al.}$ \cite{Dias.Nature2023} found two distinct hydrides in the Lu-N-H sample, both have FCC sublattices made of Lu atoms. Following this criteria, we examined the 4096 meta-stable Lu-H structures within 220 meV/atom above the convex hull. 
Four thermodynamically meta-stable structures comprising of FCC Lu-sublattice are discovered, LuH (P$\Bar{4}$3m, 89 meV/atom above the hull), LuH (F$\Bar{4}$3m, 110 meV/atom above the hull), LuH (Fm$\Bar{3}$m, 209 meV/atom above the hull) and LuH$_3$(Fm$\Bar{3}$m, 81 meV/atom above the hull), as listed in the upper panel of Fig.~\ref{hull}a by blue crosses. 
In recent theoretical studies \cite{Liu.arxiv2023, Huo.arxiv2023, Xie.CPL2023, Sun.arxiv2023} as well as in the paper by Dasenbrock-Gammon $\textit{et al.}$ \cite{Dias.Nature2023}, three of the four meta-stable structures, LuH (F$\Bar{4}$3m), LuH (Fm$\Bar{3}$m) and LuH$_3$ (Fm$\Bar{3}$m) are discussed as the potential components of Lu-N-H. 
In the Lu-N binary system at 2 GPa, two stoichiometric structures, Lu$_{10}$N$_9$ (P$\Bar{1}$) and LuN (Fm$\Bar{3}$m), are found thermodynamically stable as shown in the lower panel of Fig.~\ref{hull}a. LuN (Fm$\Bar{3}$m) is also found in several calculations \cite{Xie.CPL2023, Ferreira.arxiv2023}, and Lu$_{12}$N$_{11}$ is suggested thermodynamically stable by Xie $\textit{et al.}$ \cite{Xie.CPL2023}. 
Actually, both Lu$_{10}$N$_9$ and Lu$_{12}$N$_{11}$ could be derived from LuN (Fm$\Bar{3}$m) by removing a nitrogen-atom.
Thermodynamically stable LuH$_2$ (Fm$\Bar{3}$m) and LuN (Fm$\Bar{3}$m) processes FCC Lu-sublattice as well. Therefore, we found totally six candidate structures with FCC Lu-sublattice in our predicted structures, including LuH (P$\Bar{4}$3m), LuH (F$\Bar{4}$3m), LuH (Fm$\Bar{3}$m), LuH$_2$ (Fm$\Bar{3}$m), LuH$_3$ (Fm$\Bar{3}$m) and LuN (Fm$\Bar{3}$m). Among the six structures, LuH (Fm$\Bar{3}$m) and LuH$_3$ (Fm$\Bar{3}$m) are dynamically unstable due to the imaginary frequencies in their phonon spectra. In contrast, the phonon spectra of LuH (P$\Bar{4}$3m), LuH (F$\Bar{4}$3m), LuH$_2$ (Fm$\Bar{3}$m) and LuN (Fm$\Bar{3}$m) indicate they are dynamically stable. Please see the supplementary information for details of the crystal structures and the phonon spectra. 

 To further ascertain the compositions of Lu-N-H, the derived XRD patterns of the four dynamically stable candidate structures, LuH (P$\Bar{4}$3m), LuH (F$\Bar{4}$3m), LuH$_2$ (Fm$\Bar{3}$m) and LuN (Fm$\Bar{3}$m), are compared with the experimental XRD results of Dasenbrock-Gammon $\textit{et al.}$ \cite{Dias.Nature2023}. As presented in Fig.~\ref{hull}b, the positions of high peaks in the XRD results of the Lu-N-H sample can be perfectly reproduced by the derived XRD pattern of LuH$_2$ (Fm$\Bar{3}$m), LuH (P$\Bar{4}$3m) and LuH (F$\Bar{4}$3m), which could be the dominant components of Lu-N-H. And the positions of the low peaks can be best reproduced by LuN (Fm$\Bar{3}$m), which could be the minor component in Lu-N-H. In Dasenbrock-Gammon $\textit{et al.}$'s \cite{Dias.Nature2023} experiment, they found two types of compounds with lattice constants of 5.0289(4) Å and 4.7529(9) Å, respectively. At ambient pressure, our calculated lattice constant is 5.01675 Å for LuH$_2$ (Fm$\Bar{3}$m), 5.04109 Å for LuH (P$\Bar{4}$3m), 5.02423 Å for LuH (F$\Bar{4}$3m) and 4.75254 Å for LuN (Fm$\Bar{3}$m), all of which are very close to the experimental values. Considering the fact that the Lu-N-H sample was heated to only 65 $^{\circ}$C at 2 GPa in Dasenbrock-Gammon $\textit{et al.}$'s experiment \cite{Dias.Nature2023}, while the meta-stable LuH (P$\Bar{4}$3m) and LuH (F$\Bar{4}$3m) are at least 89 meV/atom above the hull, the likelihood that they can be synthesized in the Lu-N-H system is very small. Therefore, we argue that in Lu-N-H the dominant component is LuH$_2$ (Fm$\Bar{3}$m), together with minor LuN (Fm$\Bar{3}$m), whose crystal structures are presented in Fig.~\ref{hull}c. 

\subsection{The optical reflectivity}
\begin{figure*}[htbp]
    \centering 
    \includegraphics[width=1.0\textwidth]{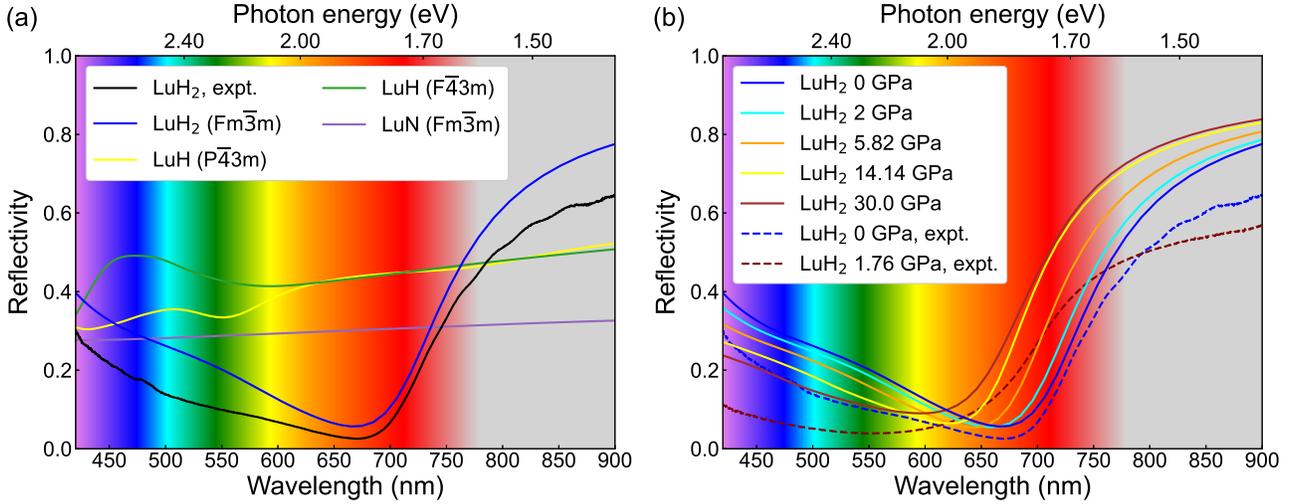}
    \caption{(Color online) (a) The optical reflectivity of LuH$_2$ (Fm$\Bar{3}$m), LuH (P$\Bar{4}$3m), LuH (F$\Bar{4}$3m) and LuN (Fm$\Bar{3}$m) at the ambient pressure. (b) The optical reflectivity of LuH$_2$ (Fm$\Bar{3}$m) under compression. The experimental results of optical reflectivity are cited from Zhao $\textit{et al.}$ \cite{Zhao.2023}.
    }
     \label{reflect}
\end{figure*}

Interesting color changes under compression are observed in the experiments of Lu-N-H \cite{Dias.Nature2023, Zhang.arxiv2023, Xing.arxiv2023} and LuH$_2$ \cite{Jin.Nature2023, Shan.CPL2023, Ming.arxiv2023}. Dasenbrock-Gammon $\textit{et al.}$ \cite{Dias.Nature2023} found that the blue Lu-N-H sample at ambient pressure turns into pink color at 0.3 GPa, then into red color at 3.2 GPa. But Cai $\textit{et al.}$ \cite{Cai.arxiv2023} found their Lu-N-H sample, which was prepared using the same method as Dasenbrock-Gammon $\textit{et al.}$, stays blue up to 6 GPa. Similar blue to red color changes were observed in LuH$_2$ by Shan $\textit{et al.}$ \cite{Shan.CPL2023}, Zhao $\textit{et al.}$ \cite{Zhao.2023} and Ming $\textit{et al.}$ \cite{Ming.arxiv2023}, whose samples turn into red color at above 3.98 GPa, 3.11 GPa and 2.5 GPa, respectively. In LuH$_{2\pm x}$N$_y$, Zhang $\textit{et al.}$ \cite{Zhang.arxiv2023} and Xing $\textit{et al.}$ \cite{Xing.arxiv2023} found an intermediate purple phase at typically 12 GPa and above, between the blue phase at ambient pressure and the red phase at higher pressure. Such purple phase was also observed in the experimental study of LuH$_2$ by Shan $\textit{et al.}$ \cite{Shan.CPL2023} and Zhao $\textit{et al.}$ \cite{Zhao.2023} at 2.23 GPa and 1.76 GPa respectively, between the blue phase and the red phase.

In order to explain the intriguing color changes in these experiments, we have calculated the optical reflectivity of the four dynamically stable candidate structures, LuH$_2$ (Fm$\Bar{3}$m), LuH (P$\Bar{4}$3m), LuH (F$\Bar{4}$3m) and LuN (Fm$\Bar{3}$m), at ambient pressure \cite{optics}. As presented in Fig.~\ref{reflect}a, the optical reflectivity of LuH$_2$ (Fm$\Bar{3}$m, blue line) has a minimum at around 675 nm, approximately the center of the red light region, indicating the reflection of red light is strongly suppressed. 
Its optical reflectivity also increases continuously at shorter wavelength, with relatively larger values in the blue light region, suggesting pronounced reflection of blue light and the blue color of LuH$_2$ (Fm$\Bar{3}$m) at ambient pressure. 
The optical reflectivity of LuH$_2$ is also measured by Zhao $\textit{et al.}$ \cite{Zhao.2023}. Our calculated optical reflectivity of LuH$_2$ (Fm$\Bar{3}$m, blue line) is nicely consistent with the experimental result (black line) at ambient pressure, as shown in Fig.~\ref{reflect}a. 
In contrast, the optical reflectivity of LuH (P$\Bar{4}$3m), LuH (F$\Bar{4}$3m) and LuN (Fm$\Bar{3}$m) have roughly flat distribution in the entire visible light region. This implies these three structures won't show blue color at ambient pressure, hence eliminates the possibility of their dominant presence in Lu-N-H. 

Our calculated optical reflectivity also supports color change in LuH$_2$ (Fm$\Bar{3}$m) under compression. As presented in Fig.~\ref{reflect}b, the position of optical reflectivity minimum shifts to the blue light region at higher pressure. 
At 30 GPa, the minimum of optical reflectivity of LuH$_2$ (Fm$\Bar{3}$m) moves to the yellow-green light region, leading to suppressed yellow-green light reflection and pronounced red-blue light reflection. 
In Zhao $\textit{et al.}$'s experiment \cite{Zhao.2023}, under compression the positions of their optical reflectivity minimum show similar shift towards blue light region like ours. The optical reflectivity of purple color LuH$_2$ sample at 1.76 GPa in Zhao $\textit{et al.}$'s experiment \cite{Zhao.2023} (Fig.~\ref{reflect}b, brown bash line) is close to our calculated optical reflectivity at 30 GPa (Fig.~\ref{reflect}b, brown solid line), which implies our numerical calculations also captured the blue to purple color change in LuH$_2$ (Fm$\Bar{3}$m) under compression. 
The difference of pressures for color changes between our calculations and Zhao $\textit{et al.}$'s experiment \cite{Zhao.2023} might be related to the pressure media in DAC. As proved by Xing $\textit{et al.}$ \cite{Xing.arxiv2023}, the critical pressures at which the color of LuH$_{2\pm x}$N$_y$ changes are sensitive to the pressure media being adopted. 
In the experiment of Xing $\textit{et al.}$ \cite{Xing.arxiv2023}, the purple color of their sample could persist up to 18.5 GPa. And in the experiment of Zhang $\textit{et al.}$ \cite{Zhang.arxiv2023}, their sample remains pink-purple even up to 42 GPa, higher than our predicted pressure of 30 GPa for the purple color LuH$_2$ (Fm$\Bar{3}$m).

We have also tested the effects of vacancies and nitrogen-doping on the optical reflectivity of LuH$_2$ (Fm$\Bar{3}$m). As presented in Fig. S2a of the supplementary information, the positions of the optical reflectivity minimum quickly shift to the blue light region as we removed up to four hydrogen-atoms from the 2$\times$2$\times$2 supercell of LuH$_2$ (Fm$\Bar{3}$m). 
The optical reflectivity minimum of Lu$_8$H$_{12}$ is at 475 nm, roughly the center of the blue light region, suggesting the quenched blue light reflection and the pronounced red light reflection. This suggests that the introduction of vacancies at hydrogen-sites could assist the blue to red color changes in LuH$_2$ (Fm$\Bar{3}$m). Similar vacant hydrogen-sites may widely exist in hydrides. 
Wang $\textit{et al.}$ \cite{Liu.PRL2021} found in clathrate hydrides, for example Li$_2$MgH$_{16}$, the hydrogen-atoms may transfer among the interstitial region within the Li$_2$Mg sublattice even at 250 GPa. The effects of nitrogen-doping on the optical reflectivity of LuH$_2$ (Fm$\Bar{3}$m) is shown in Fig. S2b of the supplementary information, which suggests nitrogen-doped LuH$_2$ (Fm$\Bar{3}$m) won't have blue or red color, hence not likely being the major component of Lu-N-H. 

Based on the analysis of optical reflectivity, we propose that the blue color of the Lu-N-H system at ambient pressure comes from the dominant presence of LuH$_{2}$ in the samples \cite{Dias.Nature2023, Ming.arxiv2023, Xing.arxiv2023, Zhang.arxiv2023, Cai.arxiv2023}. Under compression, the optical reflectivity minimum of LuH$_{2}$ gradually shifts from the red light region to the blue light region, which may be accompanied by vacancy formation at the hydrogen-sites, leading to blue to red color change at high pressures \cite{Shan.CPL2023, Zhao.2023}.

\subsection{Electronic structure, phonon spectrum, and $T_{\text{c}}$}
 
\begin{figure*}[tbp]
    \centering
    \includegraphics[width=0.75\textwidth]{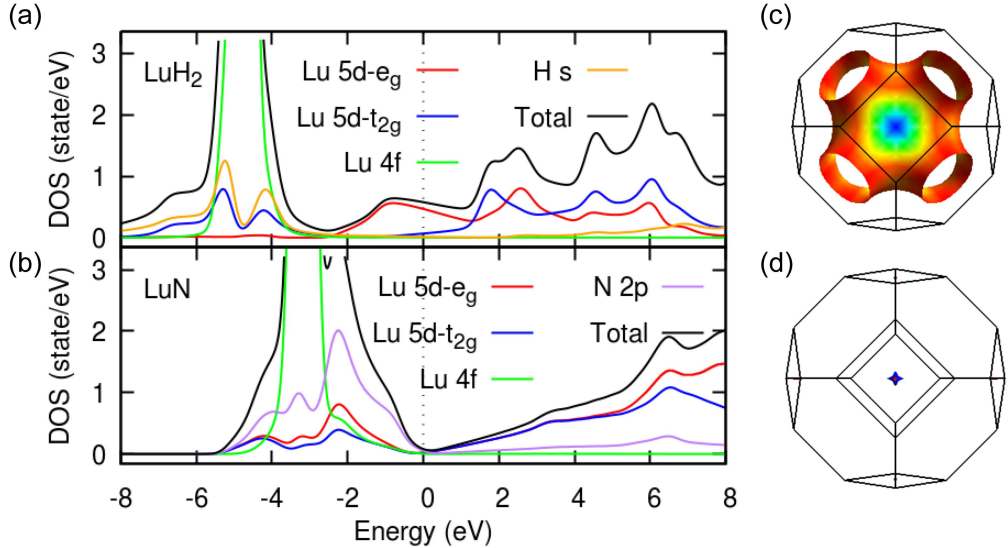}
    \caption{(Color online) The total and the partial electronic DOS and the Fermi surfaces of LuH$_2$  (Fm$\Bar{3}$m) and LuN (Fm$\Bar{3}$m) at 2 GPa.}
    \label{es}
\end{figure*}

 Our calculations of the electronic DOS of stoichiometric LuH$_2$ (Fm$\Bar{3}$m) and LuN (Fm$\Bar{3}$m) are summarized in Fig.~\ref{es}. 
 In LuH$_2$ (Fm$\Bar{3}$m) and LuN (Fm$\Bar{3}$m), the peaks of Lu-4f DOS center at $-$4.73 eV and $-$3.15 eV respectively well below the Fermi level, indicating that they are not involved in the electronic transport. 
 The DOS at the Fermi level in LuH$_2$ (Fm$\Bar{3}$m) are mainly from the Lu-5d-$e_{\text{g}}$ orbitals, consistent with experimental discovery that the electric conduction in the nitrogen-doped lutetium hydride LuH$_{2\pm x}$N$_y$ is dominated by a single band \cite{Zhang.arxiv2023}. 
 In contrast, the DOS from hydrogen in LuH$_2$ (Fm$\Bar{3}$m) is small, especially at the Fermi level. The previously discovered high-temperature superconducting rare earth hydrides share certain features, including the large electronic DOS of hydrogen at the Fermi level due to high hydrogen ratio per formula unit, and the transfer of electrons from the rare earth atoms to the hydrogen-atoms that further increases the DOS of hydrogen \cite{Peng.PRL2017}. 
 In LuH$_2$ (Fm$\Bar{3}$m) not only the DOS of hydrogen is small, there is also no sign of charge transfer between the lutetium and the hydrogen, suggesting the $T_{\text{c}}$ of stoichiometric LuH$_2$ (Fm$\Bar{3}$m) won't be large, if not be zero. Since the bandwidth of Lu-5d orbitals in LuH$_2$ (Fm$\Bar{3}$m) and LuN (Fm$\Bar{3}$m) is close to that of H-s and N-2p orbital, the correlation effect among the Lu-5d electrons is most likely weak. 
 The Fermi surfaces of LuH$_2$ (Fm$\Bar{3}$m) and LuN (Fm$\Bar{3}$m) are shown in Fig.~\ref{es}c and Fig.~\ref{es}d. The Femi surfaces of LuH$_2$ (Fm$\Bar{3}$m) resemble those of elemental Cu, which seems to suggest that the electronic structure of LuH$_2$ (Fm$\Bar{3}$m) near the Fermi level is free-electron-like. 
 In addition, the absence of any noticeable DOS peaks within 1 eV of the Fermi level suggests that LuH$_2$ (Fm$\Bar{3}$m) does not have any van Hove singularities near the Fermi level. LuN  (Fm$\Bar{3}$m) has a nearly vanishing Fermi surface due to its small DOS at the Fermi level, and is therefore unlikely to be a superconductor.

 \begin{figure*}[tbp]
    \centering
    \includegraphics[width=1.0\textwidth]{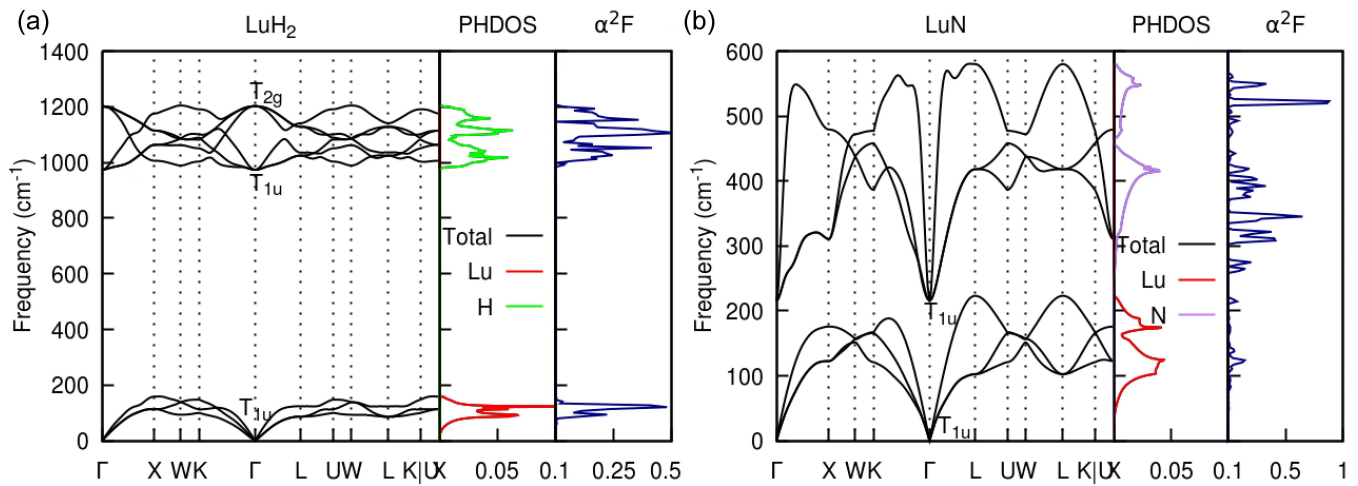}
    \caption{(Color online) The phonon dispersion relation, the total and projected phonon density of states (PHDOS), the Eliashberg function $\alpha^2$F of LuH$_2$ (Fm$\Bar{3}$m) and LuN (Fm$\Bar{3}$m) at 2 GPa.}
    \label{ph}
\end{figure*}

 Further calculations of the EPC strength and the superconducting transition temperature $T_{\text{c}}$ confirmed our conjecture. As shown in Fig.~\ref{ph}, the phonon spectra of LuH$_2$ (Fm$\Bar{3}$m) and LuN (Fm$\Bar{3}$m) indicate that these two structures are dynamically stable since there are no imaginary frequencies. 
 In LuH$_2$ (Fm$\Bar{3}$m) the PHDOS of lutetium is well separated from the PHDOS of hydrogen, but in LuN (Fm$\Bar{3}$m), the PHDOS of lutetium is closer to PHDOS of nitrogen due to relatively larger mass of nitrogen. 
 As presented in Table~\ref{epc}, the superconducting transition temperature $T_{\text{c}}$ is zero for both LuH$_2$ (Fm$\Bar{3}$m) and LuN (Fm$\Bar{3}$m). 
 Either in LuH$_2$ (Fm$\Bar{3}$m) or LuN (Fm$\Bar{3}$m), the contribution to EPC parameter $\lambda$ by the light atoms, hydrogen and nitrogen, is very small. The absence of superconductivity in these two compounds is due to the weak EPC parameter $\lambda$ and the low logarithmic frequency moment $\omega_{\text{log}}$ as shown in Table~\ref{epc}. 
 In contrast, the EPC parameter $\lambda$ and the logarithmic frequency moment $\omega_{\text{log}}$ of typical high-temperature superconducting rare earth hydrides, such as LaH$_{10}$ (Fm$\bar{3}$m), are much larger. 

 \begin{table}[t]
    \centering
    \caption{The average EPC parameter $\lambda$, the logarithmic average frequency $\omega_{\text{log}}$ and the superconducting transition temperature $T_{\text{c}}$ for LuH$_2$ (Fm$\Bar{3}$m) and LuN (Fm$\Bar{3}$m), in comparison with LaH$_{10}$ \cite{Liu.PNAS2017}.}
    \begin{tabular}{cccccccccc}
    \hline
     & Space group & P (GPa) & $\lambda$ & $\omega_{\text{log}}$ (K) & $T_{\text{c}}$ (K) & \\
    \hline
    LuH$_2$ & Fm$\bar{3}$m & 2 & 0.280 & 283.7 & 0.0 &  \\
    LuN & Fm$\bar{3}$m & 2 & 0.125 & 392.5 & 0.0 & \\
    LaH$_{10}$ & Fm$\bar{3}$m & 250 & 2.29 & 877.1 & 274-286 & \\
    \hline
    \end{tabular}
    \label{epc}
\end{table}

We have also tested the effects of nitrogen-doping on LuH$_2$ (Fm$\Bar{3}$m). In Fig. S7 of the supplementary information, the total and the hydrogen-DOS in nitrogen-doped LuH$_2$ are plotted in comparison to these of stoichiometric LuH$_2$ (Fm$\Bar{3}$m) at 2 GPa. 
As one to three hydrogen-atoms in the supercell of LuH$_2$ are replaced by nitrogen-atoms, the hydrogen-DOS at $E_{\text{F}}$ is slightly enhanced from 0.013 eV$^{-1}$ in stoichiometric LuH$_2$ (Fm$\Bar{3}$m) to maximum 0.023 eV$^{-1}$ in Lu$_8$H$_{13}$N$_3$. 
Considering the fact that the total DOS of stoichiometric LuH$_2$ (Fm$\Bar{3}$m) at $E_{\text{F}}$ is 0.6 eV$^{-1}$, 1$\times10^{-2}$ eV$^{-1}$ enhancement of hydrogen-DOS is too small. Therefore, mild nitrogen-doping would not enhance superconductivity in LuH$_2$ (Fm$\Bar{3}$m). 

\section{Conclusion}
In conclusion, we have conducted extensive numerical investigations on the Lu-H and the Lu-N binary systems, including the structure searches, the XRD pattern benchmarks, the thermodynamical stability analysis, the phonon spectra calculations, and the optical reflectivity derivations. 
Our works indicate that in the Lu-N-H system of Dasenbrock-Gammon $\textit{et al.}$  \cite{Dias.Nature2023}, the dominant component is LuH$_2$ (Fm$\bar{3}$m), together with minor LuN (Fm$\bar{3}$m). The blue LuH$_2$ (Fm$\bar{3}$m) at ambient pressure will turn into purple color and then into red color under compression, probably accompanied by the formation of vacant hydrogen-sites. 
In either LuH$_2$ (Fm$\bar{3}$m) or LuN (Fm$\bar{3}$m), the dominant electronic DOS at the Fermi level are from Lu-5d, together with very little DOS from hydrogen or nitrogen, which lead to small electron-phonon coupling and the absence of superconductivity in these two compounds. 
We also found that nitrogen-doping to LuH$_2$ (Fm$\bar{3}$m) barely increases the DOS of hydrogen, thus failing to enhance the superconductivity. 

\section{Acknowledgements}
This work is supported by the National Natural Science Foundation of China No. 11604255 and the Natural Science Basic Research Program of Shaanxi No. 2021JM-001. We would like to thank Jianjun Ying for the important discussions. The computations are performed at the TianHe-2 national supercomputing center in Guangzhou and the HPC platform of Xi’an Jiaotong University.


\end{document}


\title{Supplementary information: leading components and pressure induced color changes in the lutetium-nitrogen-hydrogen system}
\author{Xiangru Tao}
\affiliation{MOE Key Laboratory for Non-equilibrium Synthesis and Modulation of Condensed Matter, Shaanxi Province Key Laboratory of Advanced Functional Materials and Mesoscopic Physics, School of Physics, Xi'an Jiaotong University, 710049, Xi'an, Shaanxi, P.R.China}
\author{Aiqin Yang}
\affiliation{MOE Key Laboratory for Non-equilibrium Synthesis and Modulation of Condensed Matter, Shaanxi Province Key Laboratory of Advanced Functional Materials and Mesoscopic Physics, School of Physics, Xi'an Jiaotong University, 710049, Xi'an, Shaanxi, P.R.China}
\author{Shuxiang Yang}
\affiliation{Zhejiang laboratory, Hangzhou, Zhejiang, P.R.China}
\author{Yundi Quan}
\affiliation{MOE Key Laboratory for Non-equilibrium Synthesis and Modulation of Condensed Matter, Shaanxi Province Key Laboratory of Advanced Functional Materials and Mesoscopic Physics, School of Physics, Xi'an Jiaotong University, 710049, Xi'an, Shaanxi, P.R.China}
\author{Peng Zhang}
\affiliation{MOE Key Laboratory for Non-equilibrium Synthesis and Modulation of Condensed Matter, Shaanxi Province Key Laboratory of Advanced Functional Materials and Mesoscopic Physics, School of Physics, Xi'an Jiaotong University, 710049, Xi'an, Shaanxi, P.R.China}

    

\maketitle

\section{Structure searches of the Lu-H and the L-N systems}
Our USPEX structure searches have totally produced 8,123 structures, in which 4,850 structures are identified in the Lu-H system and 3,273 are identified in the Lu-N system. Among these 8,123 thermodynamically stable and meta-stable structures, 223 of them contains FCC sublattices composed by Lu-atoms. However, all these 223 structures with FCC Lu-sublattices can be derived from six structures, mostly from thermodynamically stable LuH$_2$ (Fm$\Bar{3}$m) on the convex hull, by adding or removing or distorting the hydrogen-atoms. The other five structures includes the thermodynamically stable LuN (Fm$\Bar{3}$m), as well as the thermodynamically meta-stable LuH (P$\Bar{4}$3m, 89meV/atom above the hull), LuH (F$\Bar{4}$3m, 110 meV/atom above the hull), LuH (Fm$\Bar{3}$m, 209 meV/atom above the hull) and LuH$_3$(Fm$\Bar{3}$m, 81 meV/atom above the hull).

\section{Crystal structures of stoichiometric Lu-H and Lu-N binary compounds}

In Table I, we have included the crystal structural information of all important stoichiometric Lu-H and Lu-N binary compounds in our calculations at 0 GPa and 2 GPa. 

\begin{table*}[!htbp]
  \centering
  \renewcommand{\arraystretch}{1.15}
    \setlength{\tabcolsep}{2mm}
  \caption{The crystal structural information of all important stoichiometric Lu-H and Lu-N binary compounds in our calculations.}
    \begin{tabular}{ccccccccc}
    \hline
    Formula & Space Group & 
    \makecell[c]{Lattice Constants \\ at 0 GPa (Å)} & 
    \makecell[c]{Lattice Constants \\ at 2 GPa (Å)} & 
    \makecell[c]{Chemical\\Symbol}  & 
    \makecell[c]{Wyckoff\\Site} & 
    x     & y     & z \\
    \hline
    \multirow{2}[0]{*}{LuH} & \multirow{2}[0]{*}{F$\bar{4}$3m} & \multirow{2}[0]{*}{\makecell[c]{a,b,c=5.02423\\$\alpha$,$\beta$,$\gamma$=90.00}} & \multirow{2}[0]{*}{\makecell[c]{a,b,c=4.97712\\$\alpha$,$\beta$,$\gamma$=90.00}} & Lu    & 4b    & 0.50000  & 0.00000  & 0.00000  \\
          &       &       &       & H     & 4d    & 0.25000  & 0.25000  & 0.75000  \\
   \hline
    \multirow{2}[0]{*}{LuH} & \multirow{2}[0]{*}{Fm$\bar{3}$m} & \multirow{2}[0]{*}{\makecell[c]{a,b,c=4.79064\\$\alpha$,$\beta$,$\gamma$=90.00}} & \multirow{2}[0]{*}{\makecell[c]{a,b,c=4.75097\\$\alpha$,$\beta$,$\gamma$=90.00}} & Lu    & 4b    & 0.50000  & 0.00000  & 0.00000  \\
          &       &       &       & H     & 4a    & 0.00000  & 0.00000  & 0.00000  \\
   \hline
    \multirow{2}[0]{*}{LuH$_{2}$} & \multirow{2}[0]{*}{Fm$\bar{3}$m} & \multirow{2}[0]{*}{\makecell[c]{a,b,c=5.01675\\$\alpha$,$\beta$,$\gamma$=90.00}} & \multirow{2}[0]{*}{\makecell[c]{a,b,c=4.99003\\$\alpha$,$\beta$,$\gamma$=90.00}} & Lu    & 4b    & 0.50000  & 0.00000  & 0.00000  \\
          &       &       &       & H     & 8c    & 0.25000  & 0.75000  & 0.75000  \\
   \hline
    \multirow{2}[0]{*}{LuN} & \multirow{2}[0]{*}{Fm$\bar{3}$m} & \multirow{2}[0]{*}{\makecell[c]{a,b,c=4.75254\\$\alpha$,$\beta$,$\gamma$=90.00}} & \multirow{2}[0]{*}{\makecell[c]{a,b,c=4.73609\\$\alpha$,$\beta$,$\gamma$=90.00}} & Lu    & 4b    & 0.50000  & 0.00000  & 0.00000  \\
          &       &       &       & N     & 4a    & 0.00000  & 0.00000  & 0.00000  \\
   \hline
    \multirow{3}[0]{*}{LuH$_{3}$} & \multirow{3}[0]{*}{Fm$\bar{3}$m} & \multirow{3}[0]{*}{\makecell[c]{a,b,c=4.98901\\$\alpha$,$\beta$,$\gamma$=90.00}} & \multirow{3}[0]{*}{\makecell[c]{a,b,c=4.96816\\$\alpha$,$\beta$,$\gamma$=90.00}} & Lu    & 4a    & 0.00000  & 0.00000  & 0.00000  \\
          &       &       &       & H     & 4b    & 0.50000  & 0.00000  & 0.00000  \\
          &       &       &       & H     & 8c    & 0.25000  & 0.75000  & 0.75000  \\
   \hline
   \multirow{3}[0]{*}{LuH } & \multirow{3}[0]{*}{P$\bar{4}$3m} & \multirow{3}[0]{*}{\makecell[c]{a,b,c=5.04109\\$\alpha$,$\beta$,$\gamma$=90.00}} & \multirow{3}[0]{*}{-} 
                                & Lu    & 4e    & 0.25387  & 0.74613  & 0.25387  \\
          &       &       &       & H     & 3c    & 0.50000  & 0.50000  & 0.00000  \\
          &       &       &       & H     & 1b    & 0.50000  & 0.50000  & 0.50000  \\
   \hline
    \multirow{3}[0]{*}{LuH} & \multirow{3}[0]{*}{P$\bar{4}$3m} & \multirow{3}[0]{*}{-} & \multirow{3}[0]{*}{\makecell[c]{a,b,c=4.99080\\$\alpha$,$\beta$,$\gamma$=90.00}} & Lu    & 4e    & 0.25354  & 0.74646  & 0.25354  \\
          &       &       &       & H     & 3c    & 0.50000  & 0.50000  & 0.00000  \\
          &       &       &       & H     & 1b    & 0.50000  & 0.50000  & 0.50000  \\
   \hline
    \multirow{8}[0]{*}{LuH$_{3}$} & \multirow{8}[0]{*}{R32} & \multirow{8}[0]{*}{\makecell[c]{a,b=6.12773\\ c=19.21705\\$\alpha$,$\beta$=90.00\\ $\gamma$=120.00}} & \multirow{8}[0]{*}{-} & Lu    & 9e    & 0.00480  & 0.33333  & 0.83333  \\
          &       &       &       & Lu    & 9d    & 0.00000  & 0.66200  & 0.00000  \\
          &       &       &       & H     & 18f   & 0.29901  & 0.00044  & 0.71994  \\
          &       &       &       & H     & 6c    & 0.33333  & 0.66667  & 0.85464  \\
          &       &       &       & H     & 18f   & 0.98598  & 0.69053  & 0.88709  \\
          &       &       &       & H     & 6c    & 0.66667  & 0.33333  & 0.69042  \\
          &       &       &       & H     & 3a    & 0.00000  & 0.00000  & 0.00000  \\
          &       &       &       & H     & 3b    & 0.66667  & 0.33333  & 0.83333  \\
   \hline
    \multirow{8}[0]{*}{LuH$_{3}$} & \multirow{8}[0]{*}{R32} & \multirow{8}[0]{*}{-} & \multirow{8}[0]{*}{\makecell[c]{a=6.08581\\ b=6.08581\\ c=19.01142\\$\alpha$,$\beta$=90.00\\ $\gamma$=120.00}} & Lu    & 9e    & 0.00451  & 0.33333  & 0.83333  \\
          &       &       &       & Lu    & 9d    & 0.00000  & 0.66192  & 0.00000  \\
          &       &       &       & H     & 18f   & 0.29905  & 0.00010  & 0.71970  \\
          &       &       &       & H     & 6c    & 0.33333  & 0.66667  & 0.85502  \\
          &       &       &       & H     & 18f   & 0.98669  & 0.68988  & 0.88691  \\
          &       &       &       & H     & 6c    & 0.66667  & 0.33333  & 0.69193  \\
          &       &       &       & H     & 3a    & 0.00000  & 0.00000  & 0.00000  \\
          &       &       &       & H     & 3b    & 0.66667  & 0.33333  & 0.83333  \\
   \hline
    \multirow{10}[0]{*}{Lu$_{10}$N$_{9}$} & \multirow{10}[0]{*}{P$\bar{1}$} & \multirow{10}[0]{*}{\makecell[c]{a=5.83139\\b=5.82981\\c=8.90422\\$\alpha$=102.64\\ $\beta$=109.03\\ $\gamma$=99.57}} & \multirow{10}[0]{*}{-} & Lu    & 2i    & 0.41663  & 0.95072  & 0.65572  \\
          &       &       &       & Lu    & 2i    & 0.40343  & 0.45225  & 0.15021  \\
          &       &       &       & Lu    & 2i    & 0.20003  & 0.85239  & 0.94755  \\
          &       &       &       & Lu    & 2i    & 0.79710  & 0.63626  & 0.55183  \\
          &       &       &       & Lu    & 2i    & 0.99656  & 0.74182  & 0.23981  \\
          &       &       &       & N     & 1d    & 0.00000  & 0.50000  & 0.00000  \\
          &       &       &       & N     & 2i    & 0.60196  & 0.80030  & 0.10140  \\
          &       &       &       & N     & 2i    & 0.39883  & 0.69969  & 0.40082  \\
          &       &       &       & N     & 2i    & 0.20126  & 0.09830  & 0.20167  \\
          &       &       &       & N     & 2i    & 0.19839  & 0.59956  & 0.69866  \\
   \hline
    \multirow{10}[0]{*}{Lu$_{10}$N$_{9}$} & \multirow{10}[0]{*}{P$\bar{1}$} & \multirow{10}[0]{*}{-} & \multirow{10}[0]{*}{\makecell[c]{a=5.80777\\b=5.80613\\c=8.86777\\$\alpha$=102.63\\$\beta$=109.04\\ $\gamma$=99.57}} & Lu    & 2i    & 0.41619  & 0.95078  & 0.65560  \\
          &       &       &       & Lu    & 2i    & 0.40351  & 0.45219  & 0.15020  \\
          &       &       &       & Lu    & 2i    & 0.20002  & 0.85243  & 0.94756  \\
          &       &       &       & Lu    & 2i    & 0.79722  & 0.63656  & 0.55176  \\
          &       &       &       & Lu    & 2i    & 0.99663  & 0.74196  & 0.24003  \\
          &       &       &       & N     & 1d    & 0.00000  & 0.50000  & 0.00000  \\
          &       &       &       & N     & 2i    & 0.60195  & 0.80030  & 0.10140  \\
          &       &       &       & N     & 2i    & 0.39886  & 0.69968  & 0.40081  \\
          &       &       &       & N     & 2i    & 0.20124  & 0.09830  & 0.20164  \\
          &       &       &       & N     & 2i    & 0.19842  & 0.59960  & 0.69869  \\
   \hline
    \end{tabular}%
  \label{tab:1}%
\end{table*}%

\newpage
\section{Non-stoichiometric L\lowercase{u}H$_{2}$ crystals}

We constructed 2$\times$2$\times$2 supercells of LuH$_2$ (Fm$\bar{3}$m), then 1-4 hydrogen-atoms were removed. Among all possible generated configurations, there are 1, 5, 7 and 23 nonequivalent structures in which the number of vacant-sites ranges from 1 to 4 respectively. The energy and space group of those structures are presented in Table II. We calculated the optical reflectivity of four structures with the lowest energy, including Lu$_{8}$H$_{15}$ (F$\Bar{4}$3m), Lu$_{8}$H$_{14}$ (Fd$\Bar{3}$m), Lu$_{8}$H$_{13}$ (F$\Bar{4}$3m) and Lu$_{8}$H$_{12}$ (Cmmm), as listed in Table III. 

\begin{table}[htbp]
  \centering
  \caption{The total energy and space group of LuH$_2$ supercell with 1-4 vacant hydrogen-sites at 0 GPa and 2 GPa.}
  \setlength{\tabcolsep}{25pt}
  \renewcommand{\arraystretch}{1.4}
    \begin{tabular}{ccccc}
    \hline
        \makecell[c]{Numbers\\of vacancies} & 
        Formula & 
        \makecell[c]{Energy \\ at 0 GPa (eV)} & 
        \makecell[c]{Energy \\ at 2 GPa (eV)} & 
        Space group \\
    \hline
    1     & Lu$_{8}$H$_{1}$$_{5}$ & -101.351835  & -101.311126  & F$\bar{4}$3m \\
    2     & Lu$_{8}$H$_{1}$$_{4}$ & -97.113085  & -97.100979  & Fd$\bar{3}$m \\
    2     & Lu$_{8}$H$_{1}$$_{4}$ & -96.879866  & -96.846552  & P$\bar{4}$3m \\
    2     & Lu$_{8}$H$_{1}$$_{4}$ & -96.747301  & -96.710927  & Fd$\bar{3}$m \\
    2     & Lu$_{8}$H$_{1}$$_{4}$ & -96.706995  & -96.679011  & Immm \\
    2     & Lu$_{8}$H$_{1}$$_{4}$ & -96.608309  & -96.573871  & Amm2 \\
    3     & Lu$_{8}$H$_{1}$$_{3}$ & -92.481387  & -92.452545  & F$\bar{4}$3m \\
    3     & Lu$_{8}$H$_{1}$$_{3}$ & -92.397169  & -92.373035  & Imm2 \\
    3     & Lu$_{8}$H$_{1}$$_{3}$ & -92.224395  & -92.178925  & I$\bar{4}$m2 \\
    3     & Lu$_{8}$H$_{1}$$_{3}$ & -92.111640  & -92.072793  & I$\bar{4}$m2 \\
    3     & Lu$_{8}$H$_{1}$$_{3}$ & -92.065543  & -92.017778  & Cm \\
    3     & Lu$_{8}$H$_{1}$$_{3}$ & -92.052533  & -92.014071  & Imm2 \\
    3     & Lu$_{8}$H$_{1}$$_{3}$ & -91.870654  & -91.826605  & R3m \\
    4     & Lu$_{8}$H$_{1}$$_{2}$ & -88.199538  & -88.189174  & Cmmm \\
    4     & Lu$_{8}$H$_{1}$$_{2}$ & -87.979146  & -87.934945  & Pn$\bar{3}$m \\
    4     & Lu$_{8}$H$_{1}$$_{2}$ & -87.947819  & -87.927392  & I$\bar{4}$m2 \\
    4     & Lu$_{8}$H$_{1}$$_{2}$ & -87.876788  & -87.852964  & Imm2 \\
    4     & Lu$_{8}$H$_{1}$$_{2}$ & -87.794910  & -87.769300  & C2 \\
    4     & Lu$_{8}$H$_{1}$$_{2}$ & -87.788468  & -87.758198  & Imm2 \\
    4     & Lu$_{8}$H$_{1}$$_{2}$ & -87.757929  & -87.728394  & I4$_1$md \\
    4     & Lu$_{8}$H$_{1}$$_{2}$ & -87.733044  & -87.680029  & P4$_2$/mcm \\
    4     & Lu$_{8}$H$_{1}$$_{2}$ & -87.700322  & -87.671340  & Cm \\
    4     & Lu$_{8}$H$_{1}$$_{2}$ & -87.612816  & -87.580152  & P$\bar{4}$m2 \\
    4     & Lu$_{8}$H$_{1}$$_{2}$ & -87.576614  & -87.530652  & C2 \\
    4     & Lu$_{8}$H$_{1}$$_{2}$ & -87.547372  & -87.497786  & Ima2 \\
    4     & Lu$_{8}$H$_{1}$$_{2}$ & -87.532582  & -87.498956  & I$\bar{4}$m2 \\
    4     & Lu$_{8}$H$_{1}$$_{2}$ & -87.530986  & -87.477735  & I4mm \\
    4     & Lu$_{8}$H$_{1}$$_{2}$ & -87.493821  & -87.466620  & Cmmm \\
    4     & Lu$_{8}$H$_{1}$$_{2}$ & -87.475447  & -87.427889  & P2$_1$/m \\
    4     & Lu$_{8}$H$_{1}$$_{2}$ & -87.463361  & -87.417543  & C2 \\
    4     & Lu$_{8}$H$_{1}$$_{2}$ & -87.394863  & -87.342452  & R3m \\
    4     & Lu$_{8}$H$_{1}$$_{2}$ & -87.383697  & -87.335008  & Cm \\
    4     & Lu$_{8}$H$_{1}$$_{2}$ & -87.368867  & -87.320622  & R3m \\
    4     & Lu$_{8}$H$_{1}$$_{2}$ & -87.351319  & -87.308116  & Ima2 \\
    4     & Lu$_{8}$H$_{1}$$_{2}$ & -87.130020  & -87.085958  & R3m \\
    4     & Lu$_{8}$H$_{1}$$_{2}$ & -87.118541  & -87.070297  & F$\bar{4}$3m \\
    \hline
    \end{tabular}%
  \label{tab:addlabel}%
\end{table}%

\newpage
\begin{table}[!htbp]
\setlength{\tabcolsep}{3pt}
  \renewcommand{\arraystretch}{1.4}
  \centering
  \caption{The crystal structure information of LuH$_2$ supercell with 1-4 vacant hydrogen-sites at 0 GPa and 2 GPa.}
    \begin{tabular}{ccccccccc}
    \hline
    Formula & 
    Space Group & 
    \makecell[c]{Lattice Constants \\ at 0 GPa (Å)} & 
    \makecell[c]{Lattice Constants \\ at 2 GPa (Å)} & 
    {\makecell[c]{Chemical\\Symbol}} & 
    {\makecell[c]{Wyckoff\\Site}} & 
    x     & y     & z \\
    \hline
    \multirow{7}[0]{*}{Lu$_{8}$H$_{15}$} & \multirow{7}[0]{*}{F$\bar{4}$3m} & \multirow{7}[0]{*}{\makecell[c]{a,b,c=10.02827\\$\alpha$,$\beta$,$\gamma$=90.00}} & \multirow{7}[0]{*}{\makecell[c]{a,b,c=9.96271\\$\alpha$,$\beta$,$\gamma$=90.00}} & Lu    & 16e   & 0.625000  & 0.375000  & 0.125000  \\
          &       &       &       & Lu    & 16e   & 0.125000  & 0.125000  & 0.375000  \\
          &       &       &       & H     & 24f   & 0.000000  & 0.250000  & 0.500000  \\
          &       &       &       & H     & 4c    & 0.750000  & 0.250000  & 0.750000  \\
          &       &       &       & H     & 4a    & 0.500000  & 0.500000  & 0.000000  \\
          &       &       &       & H     & 24g   & 0.750000  & 0.500000  & 0.250000  \\
          &       &       &       & H     & 4b    & 0.500000  & 0.500000  & 0.500000  \\
    \hline
    \multirow{4}[0]{*}{Lu$_{8}$H$_{14}$} & \multirow{4}[0]{*}{Fd$\bar{3}$m} & \multirow{4}[0]{*}{\makecell[c]{a,b,c=10.02827\\$\alpha$,$\beta$,$\gamma$=90.00}} & \multirow{4}[0]{*}{\makecell[c]{a,b,c=9.96271\\$\alpha$,$\beta$,$\gamma$=90.00}} & Lu    & 16d   & 0.125000  & 0.625000  & 0.125000  \\
          &       &       &       & Lu    & 16c   & 0.625000  & 0.875000  & 0.375000  \\
          &       &       &       & H     & 48f   & 0.500000  & 0.000000  & 0.250000  \\
          &       &       &       & H     & 8a    & 0.250000  & 0.250000  & 0.250000  \\
    \hline
    \multirow{5}[0]{*}{Lu$_{8}$H$_{13}$} & \multirow{5}[0]{*}{F$\bar{4}$3m} & \multirow{5}[0]{*}{\makecell[c]{a,b,c=10.02827\\$\alpha$,$\beta$,$\gamma$=90.00}} & \multirow{5}[0]{*}{\makecell[c]{a,b,c=9.96271\\$\alpha$,$\beta$,$\gamma$=90.00}} & Lu    & 16e   & 0.625000  & 0.375000  & 0.125000  \\
          &       &       &       & Lu    & 16e   & 0.125000  & 0.125000  & 0.375000  \\
          &       &       &       & H     & 24f   & 0.000000  & 0.250000  & 0.500000  \\
          &       &       &       & H     & 24g   & 0.750000  & 0.500000  & 0.250000  \\
          &       &       &       & H     & 4b    & 0.500000  & 0.500000  & 0.500000  \\
    \hline
    \multirow{5}[0]{*}{Lu$_{8}$H$_{12}$} & \multirow{5}[0]{*}{Cmmm} & \multirow{5}[0]{*}{\makecell[c]{a=7.09106\\ b=10.02827\\ c=3.54553\\$\alpha$,$\beta$,$\gamma$=90.00}} & \multirow{5}[0]{*}{\makecell[c]{a=7.04470\\b=9.96271\\ c=3.52235\\$\alpha$,$\beta$,$\gamma$=90.00}} & Lu    & 2b    & 0.000000  & 0.500000  & 0.000000  \\
          &       &       &       & Lu    & 4f    & 0.750000  & 0.750000  & 0.500000  \\
          &       &       &       & Lu    & 2a    & 0.000000  & 0.000000  & 0.000000  \\
          &       &       &       & H     & 8p    & 0.750000  & 0.875000  & 0.000000  \\
          &       &       &       & H     & 4j    & 0.000000  & 0.125000  & 0.500000  \\
    \hline
    \end{tabular}%
  \label{tab:addlabel}%
\end{table}%

\hspace{1cm}\\
We also randomly replaced 1-3 hydrogen-atoms in above LuH$_2$ supercells by nitrogen-atoms, whose structure information are provided in Table IV.

\begin{table}[!htbp]
\setlength{\tabcolsep}{5pt}
  \renewcommand{\arraystretch}{1.1}
  \centering
  \caption{The crystal structure information of LuH$_2$ supercell with nitrogen-doping at 2 GPa.}
    \begin{tabular}{ccccccccc}
    \hline
    Formula & 
    Space Group & 
    \makecell[c]{Lattice Constants (Å)} &  
    {\makecell[c]{Chemical\\Symbol}} & 
    {\makecell[c]{Wyckoff\\Site}} & 
    x     & y     & z \\
    \hline
    \multirow{10}[0]{*}{Lu$_{8}$H$_{13}$N$_{3}$} & \multirow{10}[0]{*}{I$\overline{4}$m2} & \multirow{10}[0]{*}{\makecell[c]{a,b=7.04470\\c=9.96271\\$\alpha$,$\beta$,$\gamma$=90.00}} & Lu    & 8i    & 0.750000  & 0.000000  & 0.125000  \\
          &       &       & Lu    & 8i    & 0.250000  & 0.000000  & 0.625000  \\
          &       &       & H     & 8h    & 0.750000  & 0.250000  & 0.250000  \\
          &       &       & H     & 2a    & 0.000000  & 0.000000  & 0.000000  \\
          &       &       & H     & 2b    & 0.000000  & 0.000000  & 0.500000  \\
          &       &       & H     & 8g    & 0.250000  & 0.250000  & 0.000000  \\
          &       &       & H     & 2c    & 0.000000  & 0.500000  & 0.250000  \\
          &       &       & H     & 4e    & 0.000000  & 0.000000  & 0.250000  \\
          &       &       & N     & 4f    & 0.000000  & 0.500000  & 0.000000  \\
          &       &       & N     & 2d    & 0.500000  & 0.000000  & 0.250000  \\
    \hline
    \multirow{14}[0]{*}{Lu$_{8}$H$_{13}$N$_{3}$} & \multirow{14}[0]{*}{Imm2} & \multirow{14}[0]{*}{\makecell[c]{a,b=7.04470\\c=9.96271\\$\alpha$,$\beta$,$\gamma$=90.00}} & Lu    & 4d    & 0.500000  & 0.250000  & 0.375000  \\
          &       &       & Lu    & 4c    & 0.250000  & 0.000000  & 0.125000  \\
          &       &       & Lu    & 4c    & 0.250000  & 0.500000  & 0.125000  \\
          &       &       & Lu    & 4d    & 0.000000  & 0.250000  & 0.375000  \\
          &       &       & H     & 2a    & 0.000000  & 0.000000  & 0.000000  \\
          &       &       & H     & 2b    & 0.000000  & 0.500000  & 0.000000  \\
          &       &       & H     & 8e    & 0.250000  & 0.250000  & 0.250000  \\
          &       &       & H     & 2a    & 0.000000  & 0.000000  & 0.250000  \\
          &       &       & H     & 2a    & 0.000000  & 0.000000  & 0.750000  \\
          &       &       & H     & 8e    & 0.250000  & 0.750000  & 0.500000  \\
          &       &       & H     & 2b    & 0.500000  & 0.000000  & 0.250000  \\
          &       &       & N     & 2b    & 0.000000  & 0.500000  & 0.250000  \\
          &       &       & N     & 2a    & 0.000000  & 0.000000  & 0.500000  \\
          &       &       & N     & 2b    & 0.500000  & 0.000000  & 0.000000  \\
    \hline
    \multirow{5}[0]{*}{Lu$_{8}$H$_{14}$N$_{2}$} & \multirow{5}[0]{*}{Fd$\overline{3}$m} & \multirow{5}[0]{*}{\makecell[c]{a,b,c=9.96271\\$\alpha$,$\beta$,$\gamma$=90.00}} & Lu    & 16d   & 0.625000  & 0.125000  & 0.125000  \\
          &       &       & Lu    & 16c   & 0.875000  & 0.875000  & 0.125000  \\
          &       &       & H     & 8b    & 0.750000  & 0.750000  & 0.750000  \\
          &       &       & H     & 48f   & 0.000000  & 0.750000  & 0.000000  \\
          &       &       & N     & 8a    & 0.000000  & 0.000000  & 0.000000  \\
    \hline
    \multirow{8}[0]{*}{Lu$_{8}$H$_{14}$N$_{2}$} & \multirow{8}[0]{*}{Immm} & \multirow{8}[0]{*}{\makecell[c]{a,b=7.04470\\c=9.96271\\$\alpha$,$\beta$,$\gamma$=90.00}} & Lu    & 4g    & 0.000000  & 0.250000  & 0.000000  \\
          &       &       & Lu    & 8m    & 0.750000  & 0.500000  & 0.250000  \\
          &       &       & Lu    & 4h    & 0.500000  & 0.250000  & 0.000000  \\
          &       &       & H     & 4j    & 0.000000  & 0.500000  & 0.125000  \\
          &       &       & H     & 4j    & 0.500000  & 0.000000  & 0.125000  \\
          &       &       & H     & 16o   & 0.250000  & 0.250000  & 0.875000  \\
          &       &       & H     & 4i    & 0.000000  & 0.000000  & 0.125000  \\
          &       &       & N     & 4i    & 0.000000  & 0.000000  & 0.375000  \\
    \hline
    \multirow{8}[0]{*}{Lu$_{8}$H$_{15}$N} & \multirow{8}[0]{*}{F$\overline{4}$3m} & \multirow{8}[0]{*}{\makecell[c]{a,b,c=9.96271\\$\alpha$,$\beta$,$\gamma$=90.00}} & Lu    & 16e   & 0.375000  & 0.125000  & 0.375000  \\
          &       &       & Lu    & 16e   & 0.625000  & 0.125000  & 0.125000  \\
          &       &       & H     & 4a    & 0.000000  & 0.000000  & 0.000000  \\
          &       &       & H     & 4b    & 0.500000  & 0.000000  & 0.000000  \\
          &       &       & H     & 24g   & 0.750000  & 0.250000  & 0.000000  \\
          &       &       & H     & 24f   & 0.500000  & 0.000000  & 0.750000  \\
          &       &       & H     & 4c    & 0.250000  & 0.250000  & 0.250000  \\
          &       &       & N     & 4d    & 0.250000  & 0.250000  & 0.750000  \\
    \hline
    \multirow{11}[0]{*}{Lu$_{8}$H$_{21}$N} & \multirow{11}[0]{*}{P$\overline{4}$m2} & \multirow{11}[0]{*}{\makecell[c]{a,b=7.17780\\c=5.19510\\$\alpha$,$\beta$,$\gamma$=90.00}} & Lu    & 4j    & 0.751000  & 0.000000  & 0.257000  \\
          &       &       & Lu    & 4k    & 0.237000  & 0.500000  & 0.230000  \\
          &       &       & H     & 4k    & 0.183000  & 0.500000  & 0.817000  \\
          &       &       & H     & 4i    & 0.740000  & 0.740000  & 0.500000  \\
          &       &       & H     & 4h    & 0.249000  & 0.249000  & 0.000000  \\
          &       &       & H     & 1a    & 0.000000  & 0.000000  & 0.000000  \\
          &       &       & H     & 2g    & 0.500000  & 0.000000  & 0.509000  \\
          &       &       & H     & 4j    & 0.338000  & 0.000000  & 0.843000  \\
          &       &       & H     & 1b    & 0.500000  & 0.500000  & 0.000000  \\
          &       &       & H     & 1c    & 0.500000  & 0.500000  & 0.500000  \\
          &       &       & N     & 1d    & 0.000000  & 0.000000  & 0.500000  \\
    \hline
    \end{tabular}%
  \label{tab:addlabel}%
\end{table}%

\hspace{1cm}
\newpage 

\section{Optical reflectivity}
 
\begin{figure*}[!htbp]
    \centering 
    \subfigure[]{
        \includegraphics[width=0.48\textwidth]{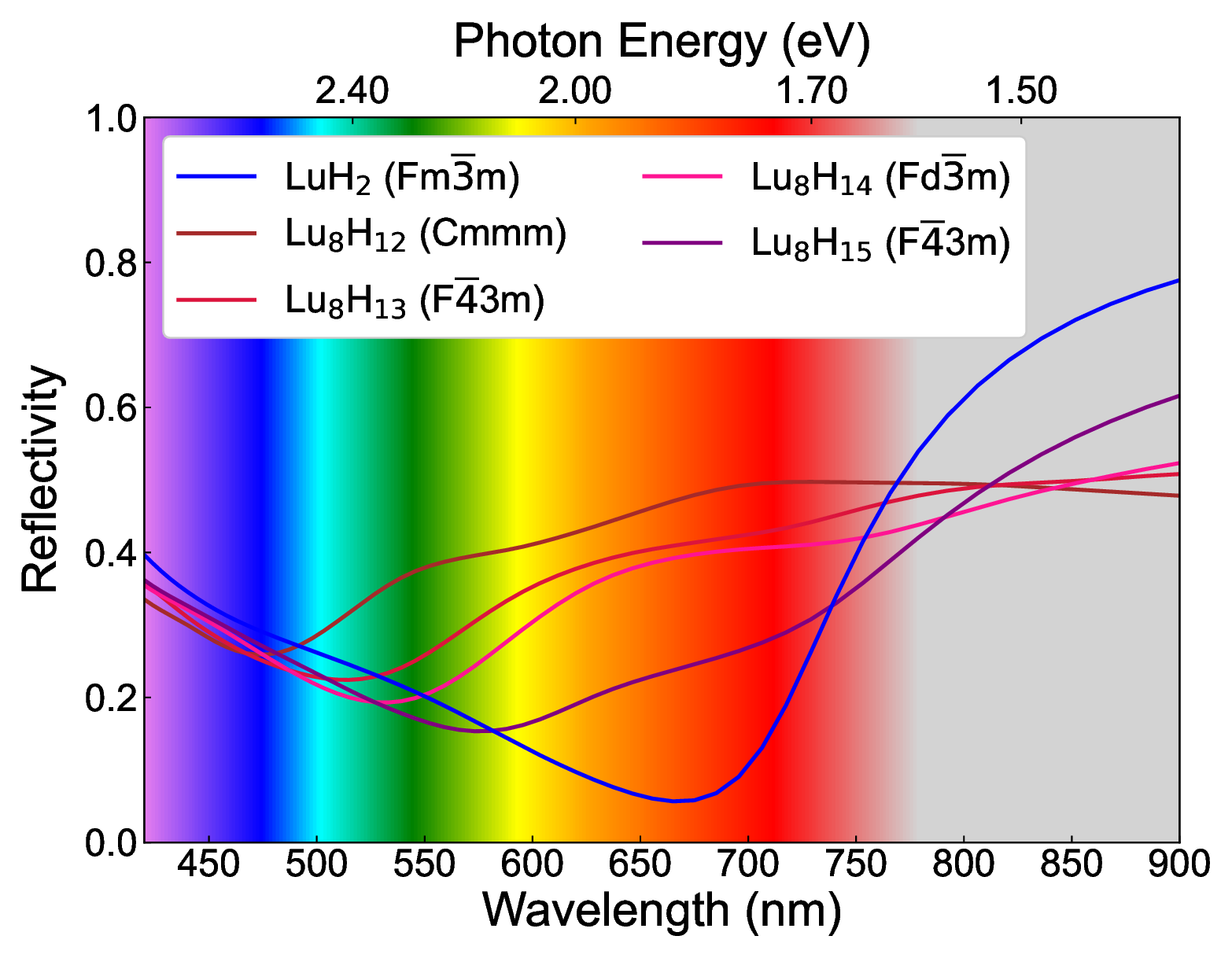}
        }
    \subfigure[]{
        \includegraphics[width=0.48\textwidth]{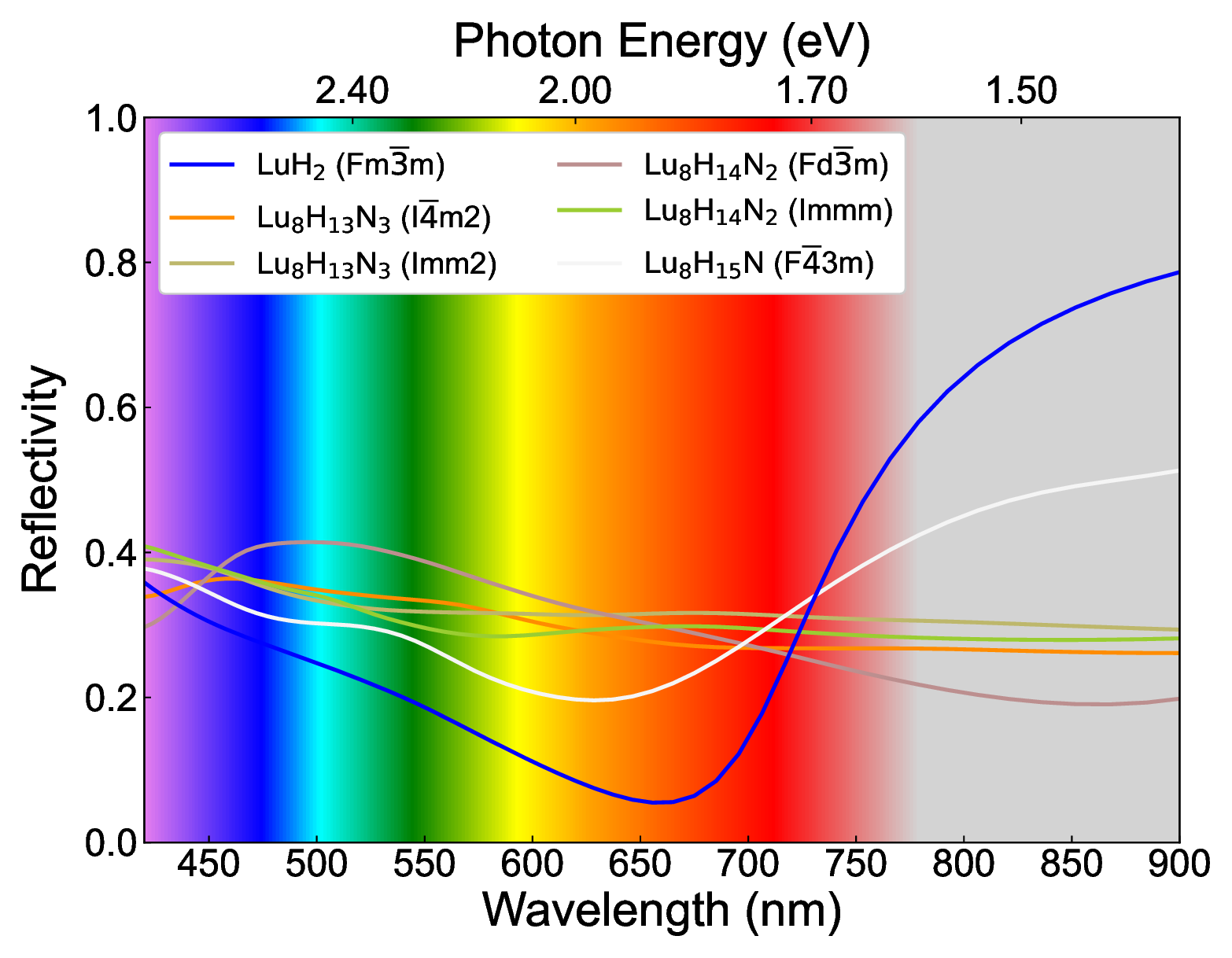}
    } 
    \caption{The calculated optical reflectivity of LuH$_2$ supercell with (a) 1-4 vacant hydrogen-sites (Lu$_8$H$_{15}$, Lu$_8$H$_{14}$, Lu$_8$H$_{13}$, Lu$_8$H$_{12}$) at the ambient pressure, and (b) 1-3 nitrogen-atoms (F$\overline{4}$3m-Lu$_8$H$_{15}$N, Fd$\overline{3}$m-Lu$_8$H$_{14}$N$_2$, Immm-Lu$_8$H$_{14}$N$_2$, I$\overline{4}$m2-Lu$_8$H$_{13}$N$_3$, Imm2-Lu$_8$H$_{13}$N$_3$) at 2 GPa.}
    \label{}
\end{figure*}

\begin{table}[!htbp]
    \centering
    \caption{The plasma frequencies in eV used in our optical reflectivity calculations.}
    \setlength{\tabcolsep}{10pt}
    \renewcommand{\arraystretch}{1.4}
    \begin{tabular}{cccccccc}
    \hline
    Formula & Space group & 0 GPa & 2 GPa & 5.82 GPa & 14.14 GPa & 30 GPa \\
    \hline
    LuH$_2$ & Fm$\Bar{3}$m & 5.03 & 5.07 & 5.18 & 5.34 & 5.58 \\
    LuH & P$\Bar{4}$3m & 2.76 & 2.76 & - & - & - \\ 
    LuH & F$\Bar{4}$3m & 3.24 & 3.27 & - & - & - \\
    LuH & Fm$\Bar{3}$m & 6.22 & 6.35 & - & - & - \\
    LuH$_3$ & Fm$\Bar{3}$m & 1.83 & 1.99 & - & - & - \\
    LuN & Fm$\Bar{3}$m & 0.29 & 0.42 & - & - & - \\
    Lu$_8$H$_{15}$ & F$\Bar{4}$3m & 3.35 & 3.41 & - & - & - \\
    Lu$_8$H$_{14}$ & Fd$\Bar{3}$m & 3.36 & 3.41 & - & - & - \\
    Lu$_8$H$_{13}$ & F$\Bar{4}$3m & 3.12 & 3.12 & - & - & - \\
    Lu$_8$H$_{12}$ & Cmmm & 3.21 & 3.26 & - & - & - \\
    Lu$_8$H$_{15}$N & F$\Bar{4}$3m & - & 2.12 & - & - & - \\
    Lu$_8$H$_{14}$N$_2$ & Fd$\Bar{3}$m & - & 3.44 & - & - & - \\
    Lu$_8$H$_{14}$N$_2$ & Immm & - & 1.91 & - & - & - \\
    Lu$_8$H$_{13}$N$_3$ & I$\Bar{4}$m2 & - & 1.42 & - & - & - \\
    Lu$_8$H$_{13}$N$_3$ & Imm2 & - & 2.11 & - & - & - \\
    \hline
    \end{tabular}
    
    \label{}
\end{table}


\newpage
\section{Electronic band structure, density of states and phonon spectra}

\begin{figure*}[!htbp]
    \centering
    \includegraphics[width=0.48\textwidth]{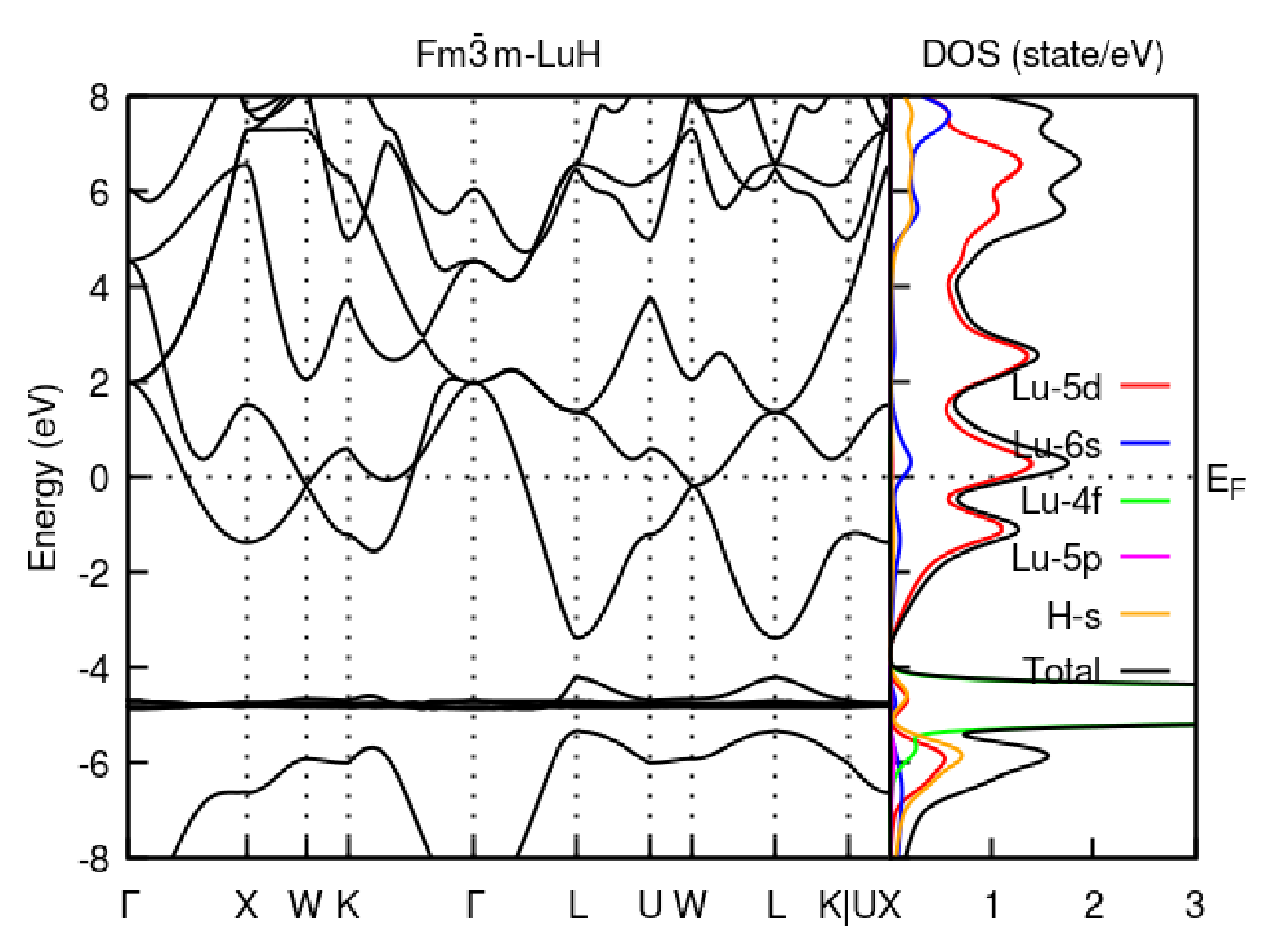}
    \includegraphics[width=0.48\textwidth]{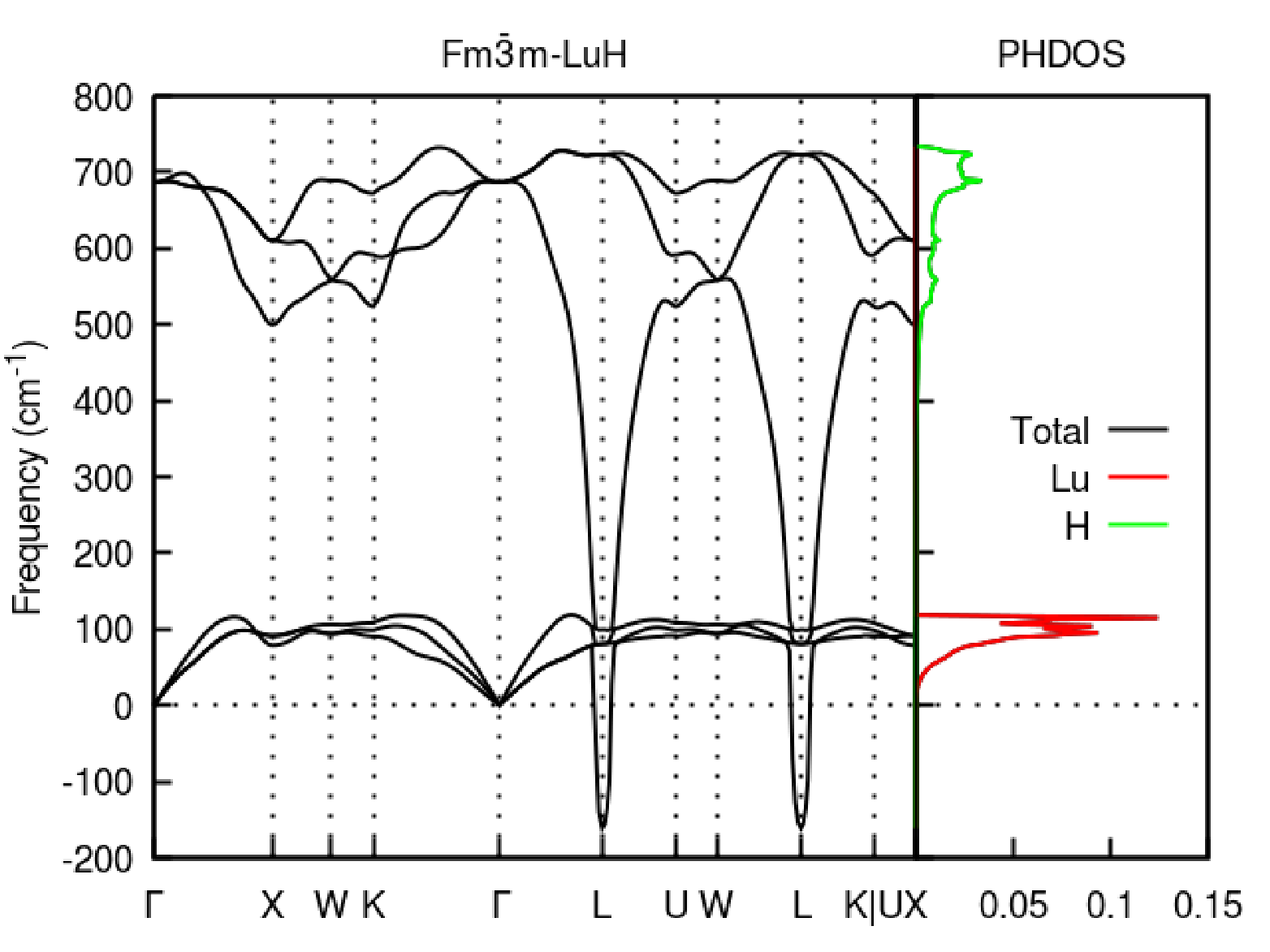}\\
    \includegraphics[width=0.48\textwidth]{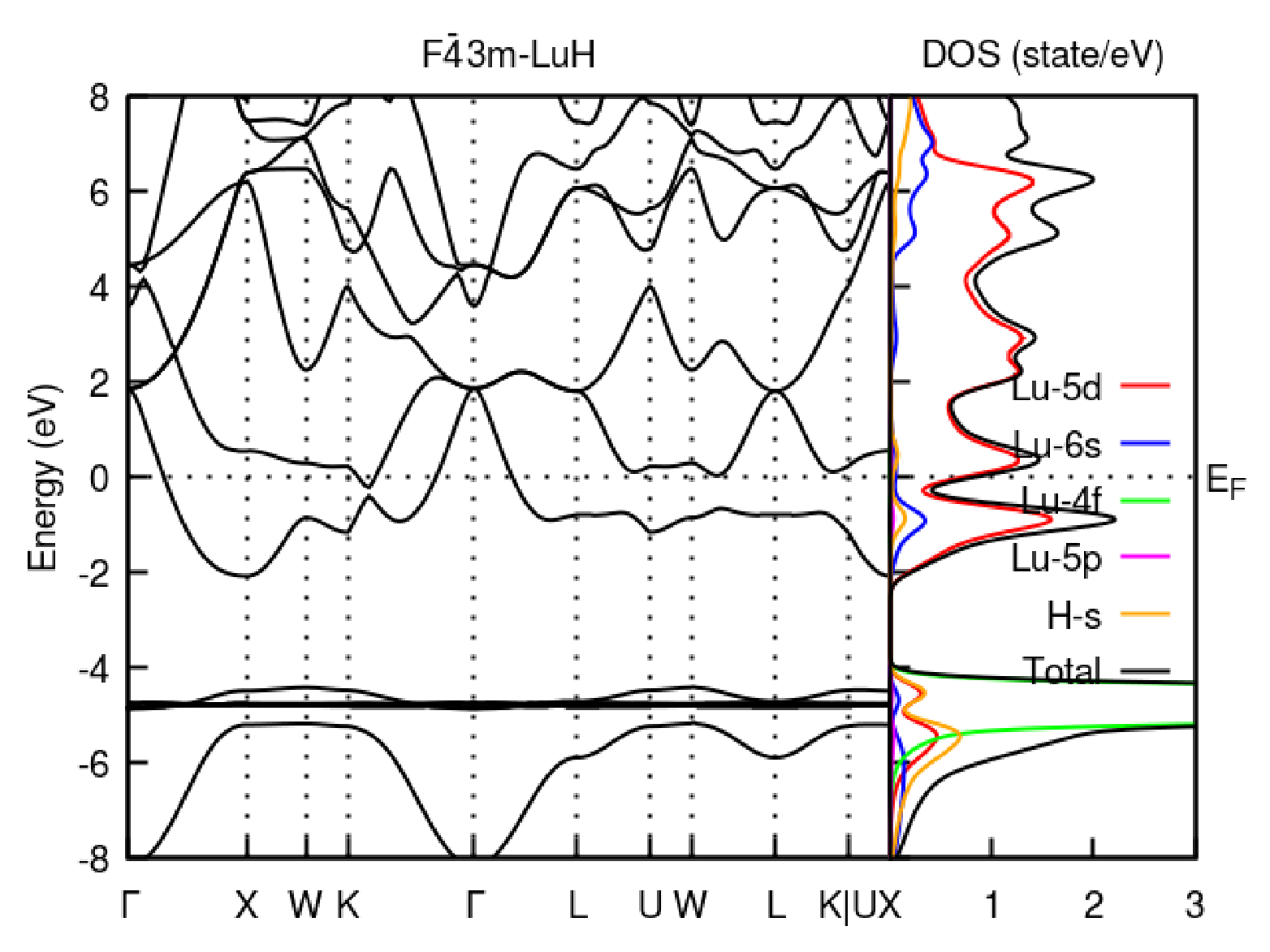}
    \includegraphics[width=0.48\textwidth]{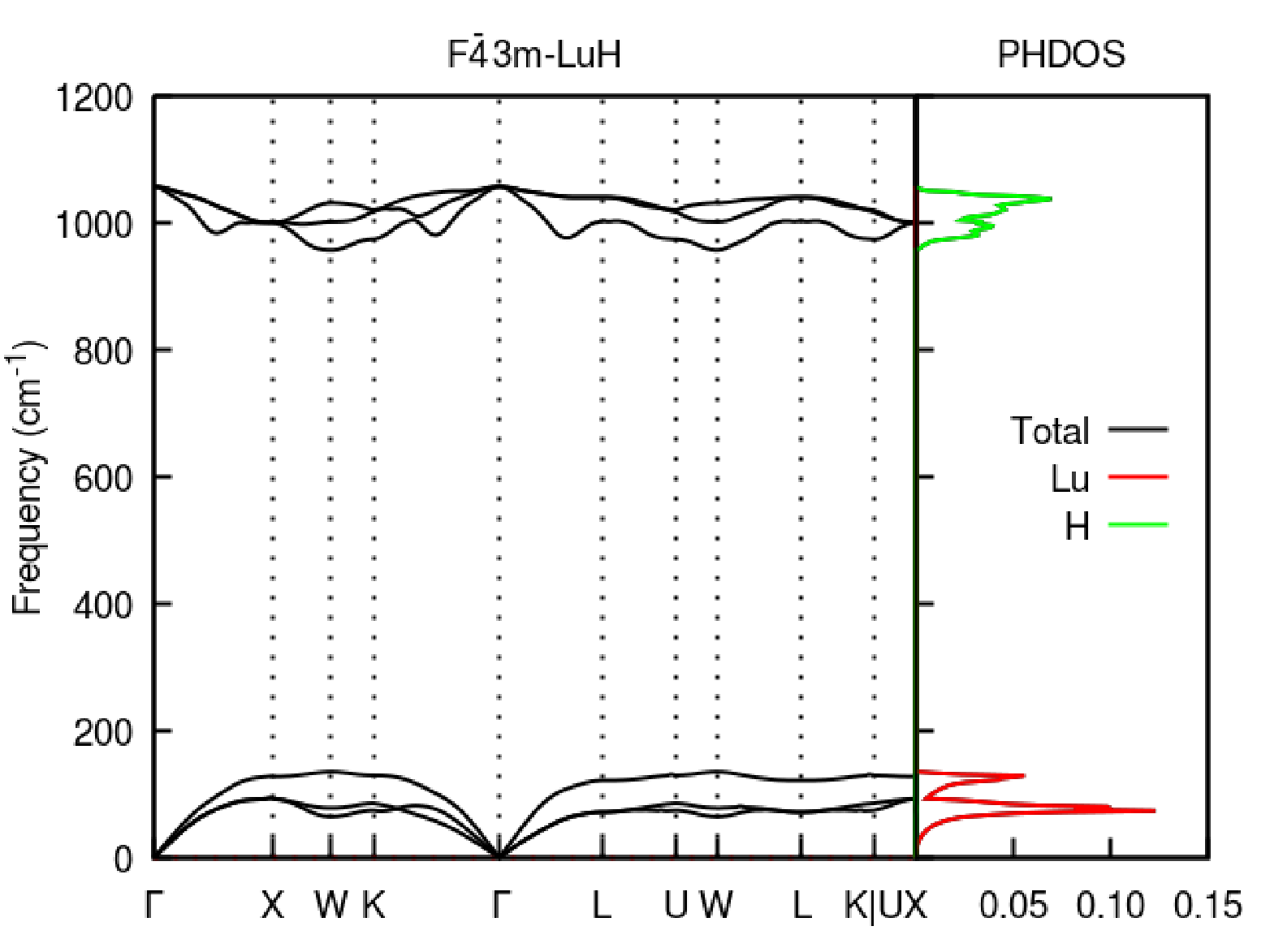}\\
    \includegraphics[width=0.48\textwidth]{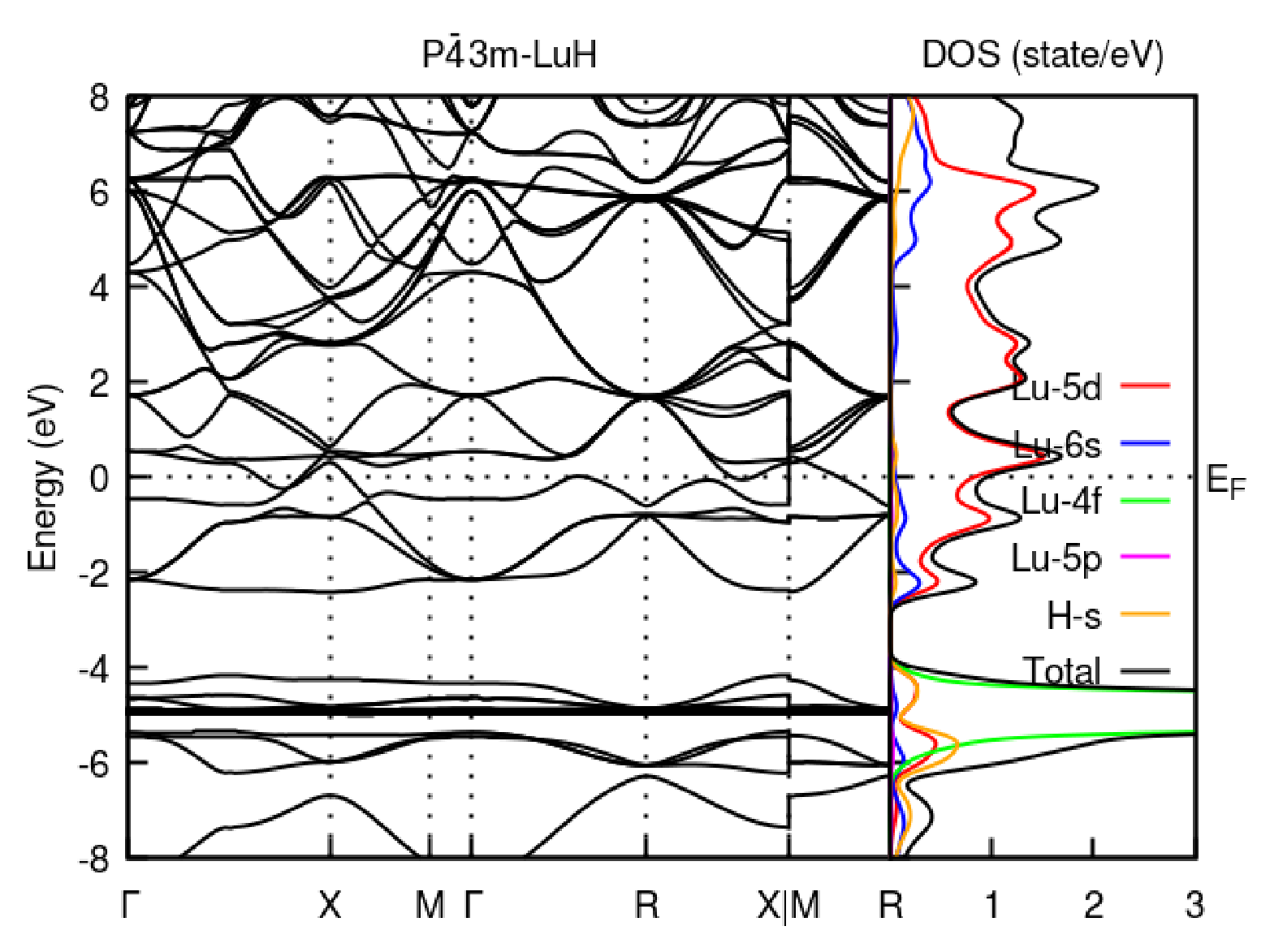}
    \includegraphics[width=0.48\textwidth]{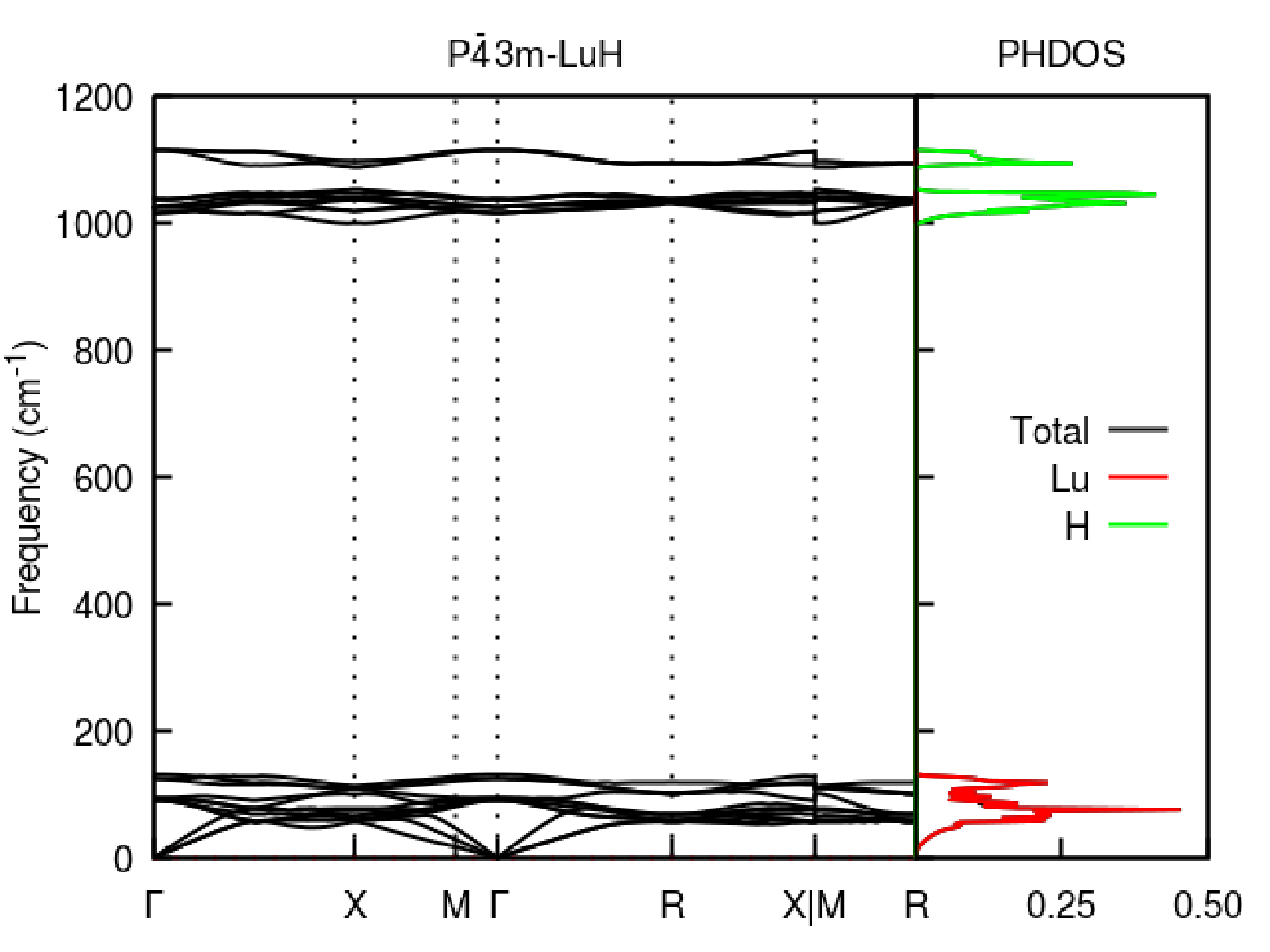}
    \caption{The electronic structure and the phonon spectra of LuH at 2 GPa.}
    \label{}
\end{figure*}

\begin{figure*}[htbp]
    \centering
    \includegraphics[width=0.47\textwidth]{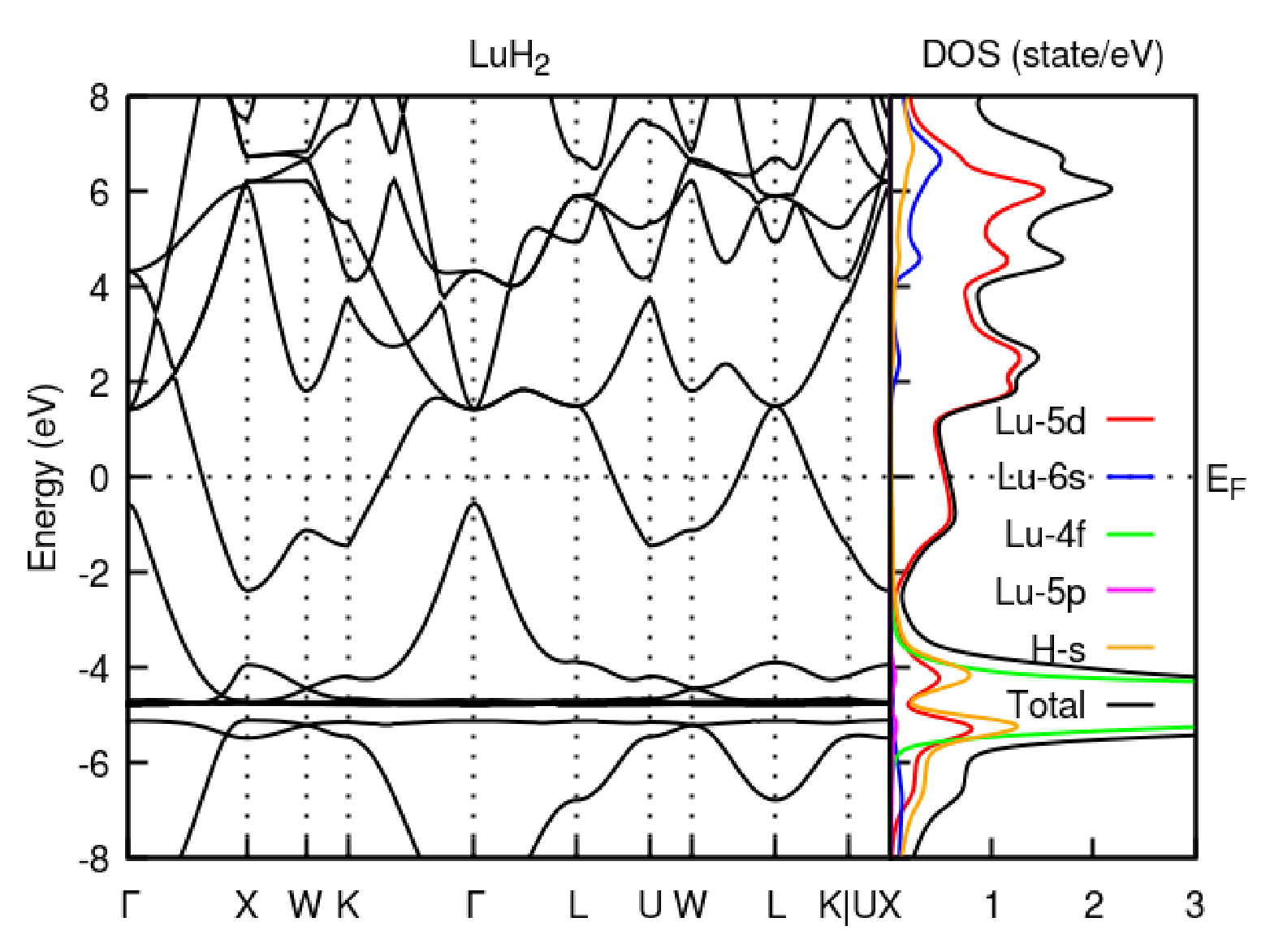}
    \includegraphics[width=0.47\textwidth]{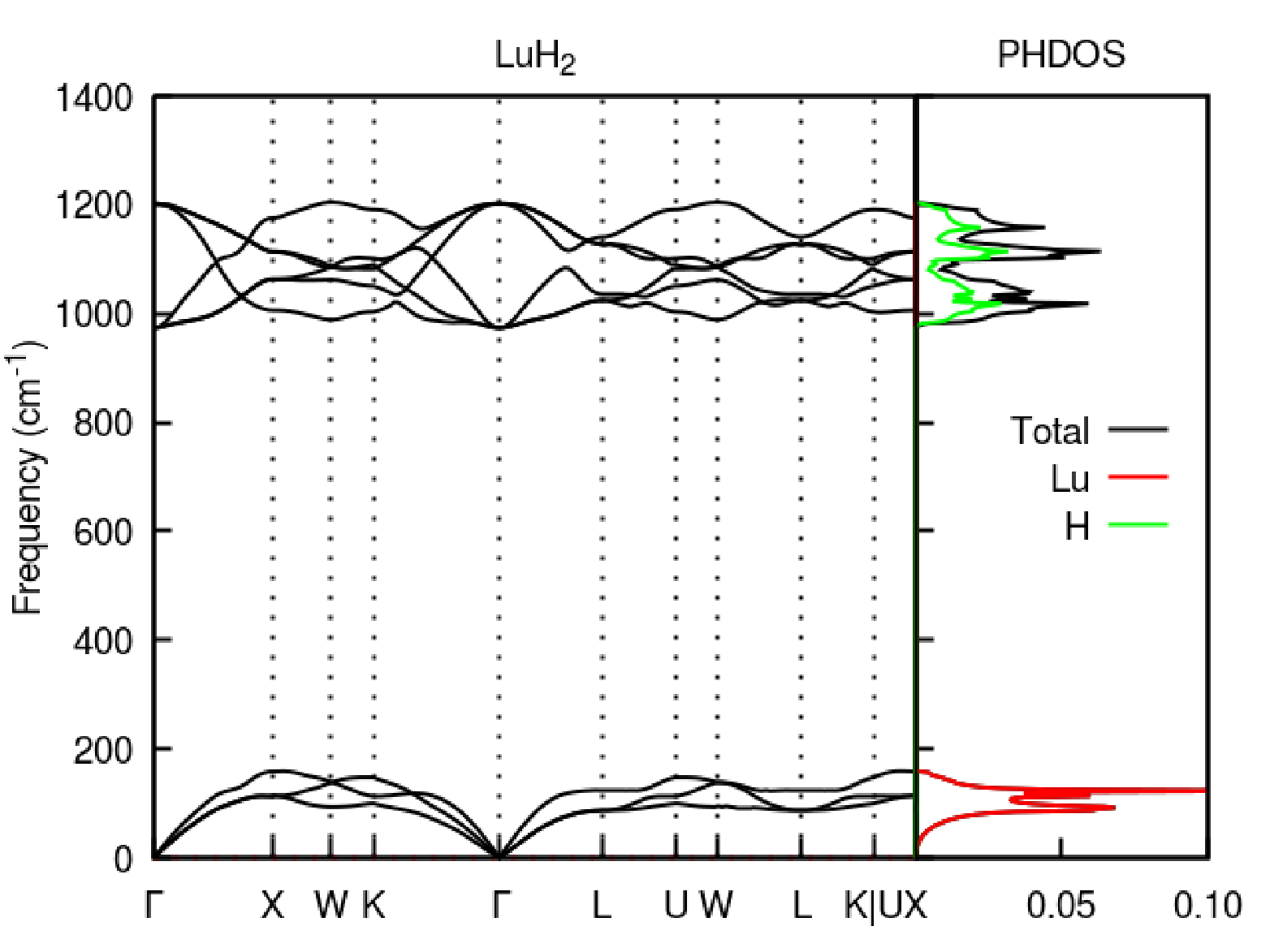}
    \caption{The electronic structure and the phonon spectrum of LuH$_2$ at 2 GPa.}
    \label{}
\end{figure*}

\begin{figure*}[!htbp]
    \centering
    \includegraphics[width=0.47\textwidth]{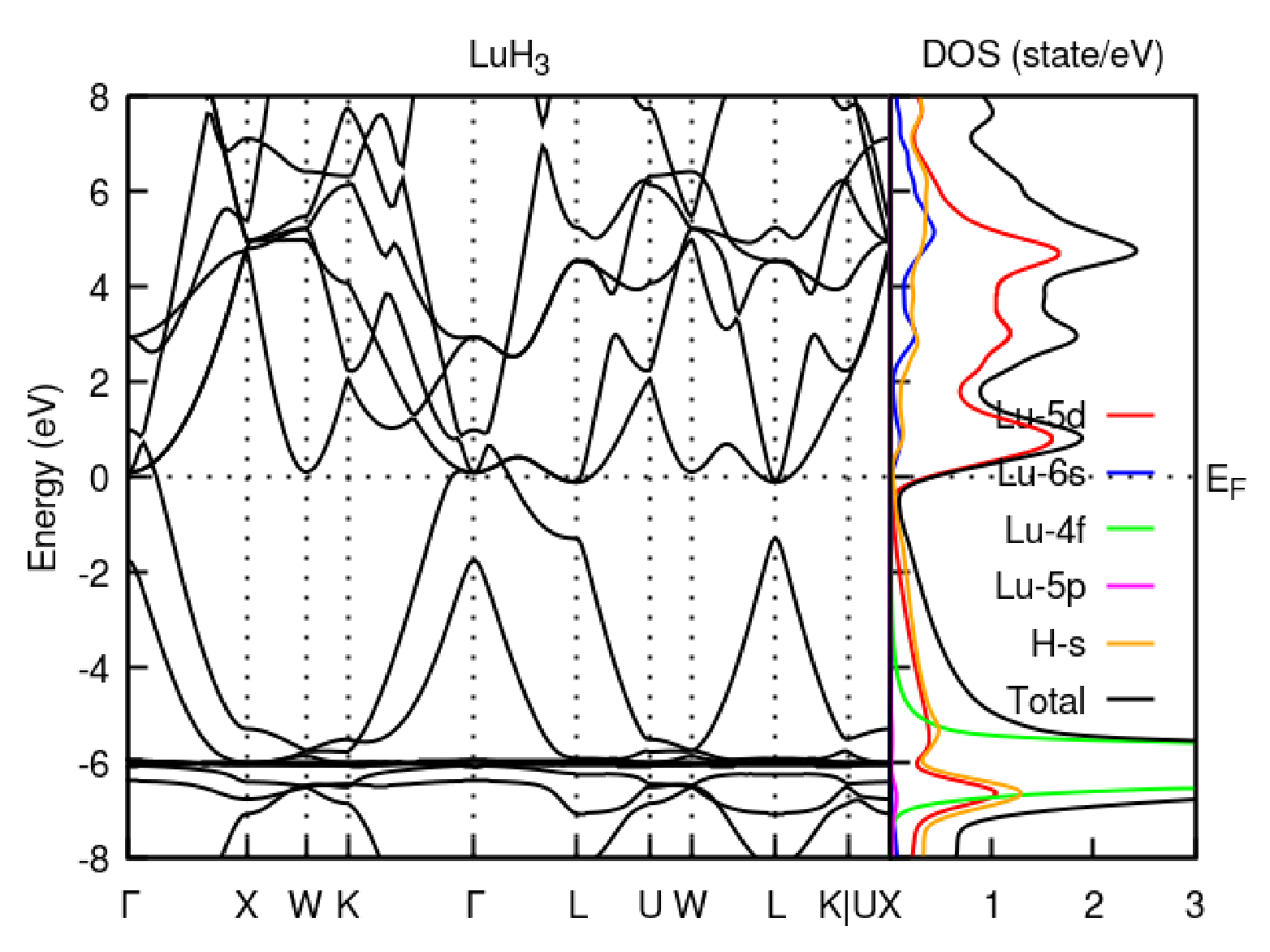}
    \includegraphics[width=0.47\textwidth]{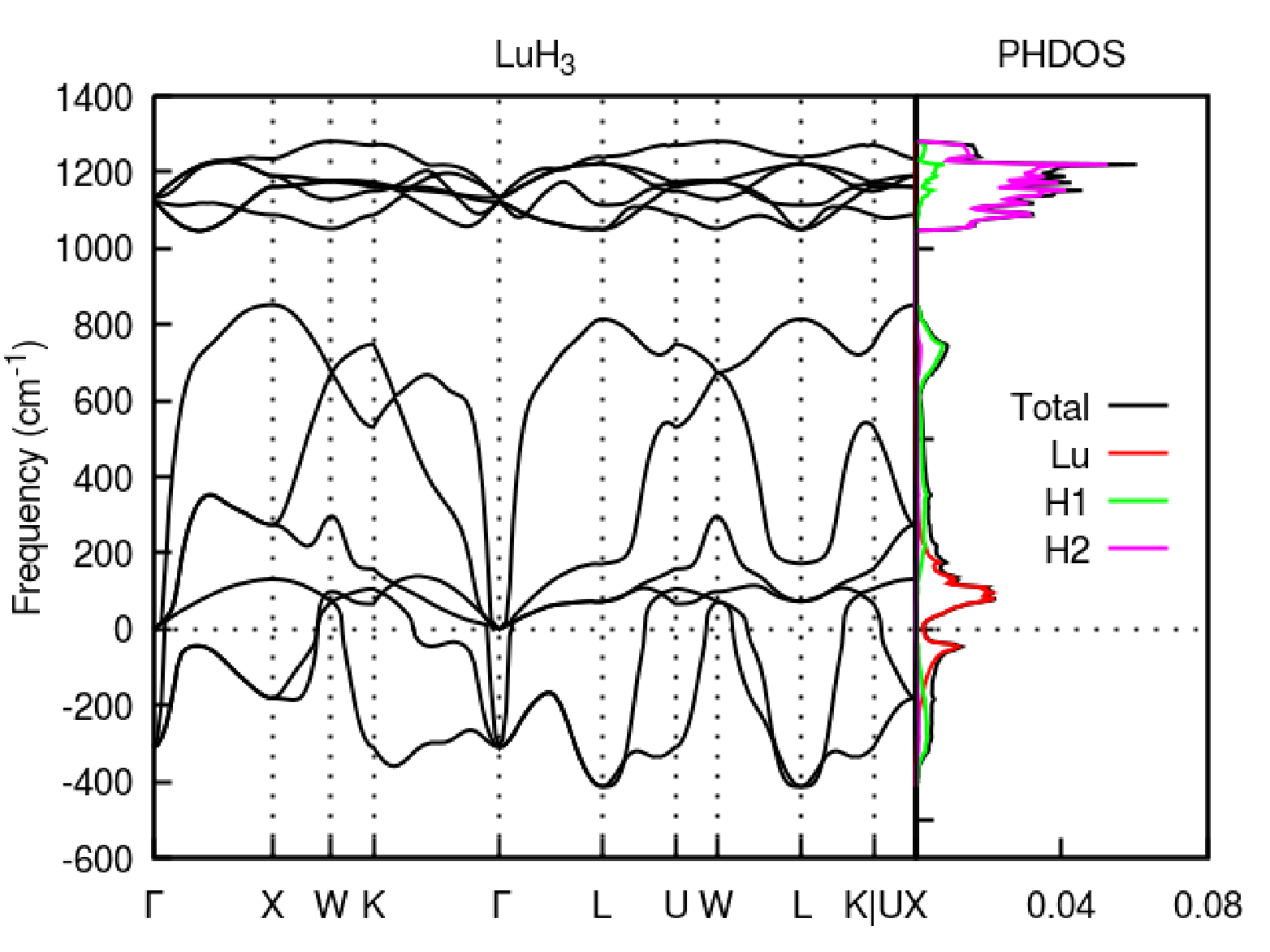}
    \caption{The electronic structure and the phonon spectrum of LuH$_3$ at 2 GPa.}
    \label{ph_3}
\end{figure*}

\begin{figure*}[!htbp]
    \centering
    \includegraphics[width=0.48\textwidth]{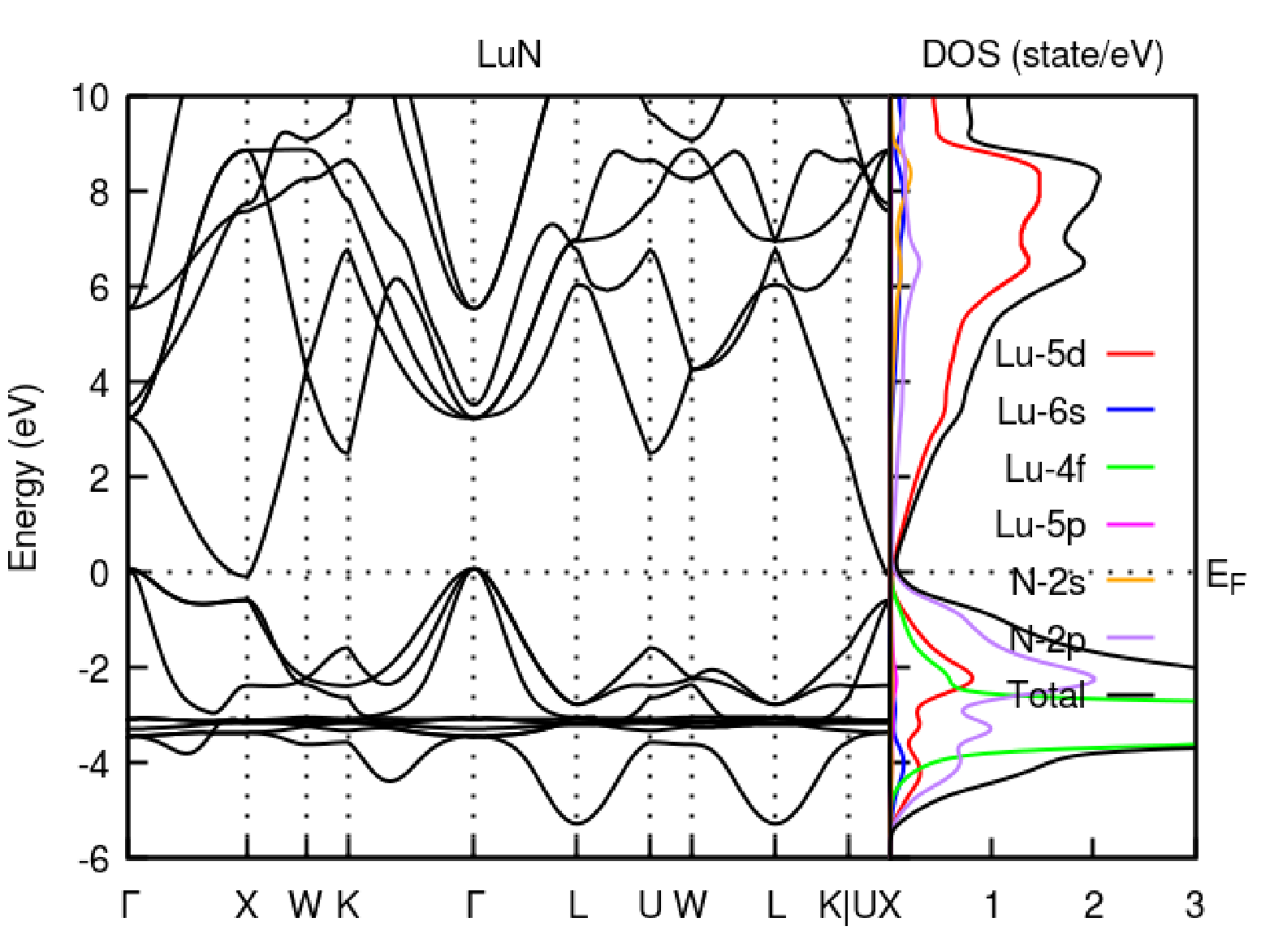}
    \includegraphics[width=0.48\textwidth]{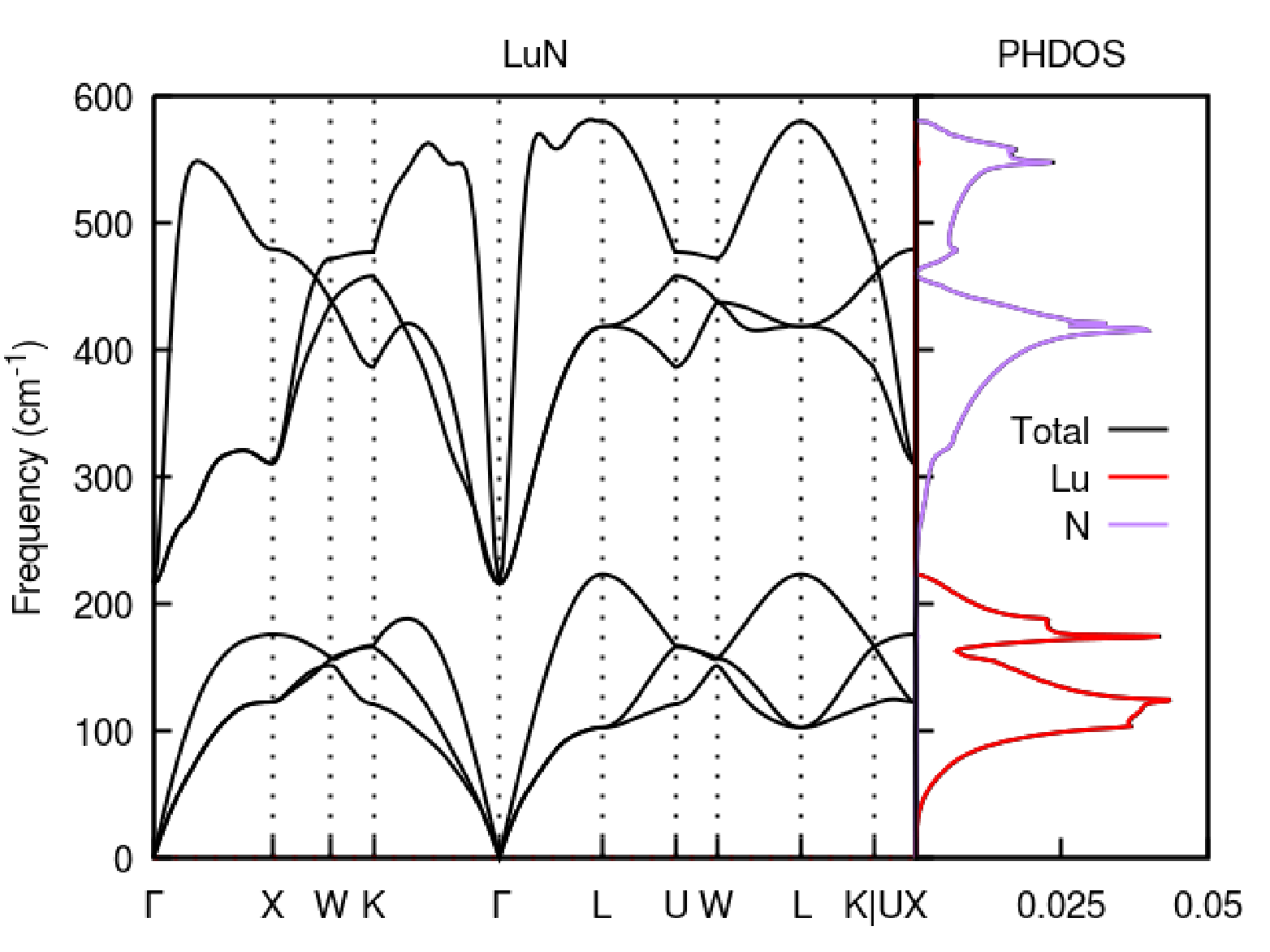}
    \caption{The electronic structure and the phonon spectrum of LuN at 2 GPa.}
    \label{}
\end{figure*}
\newpage

\begin{figure*}[!htbp]
    \centering
    \includegraphics[width=0.48\textwidth]{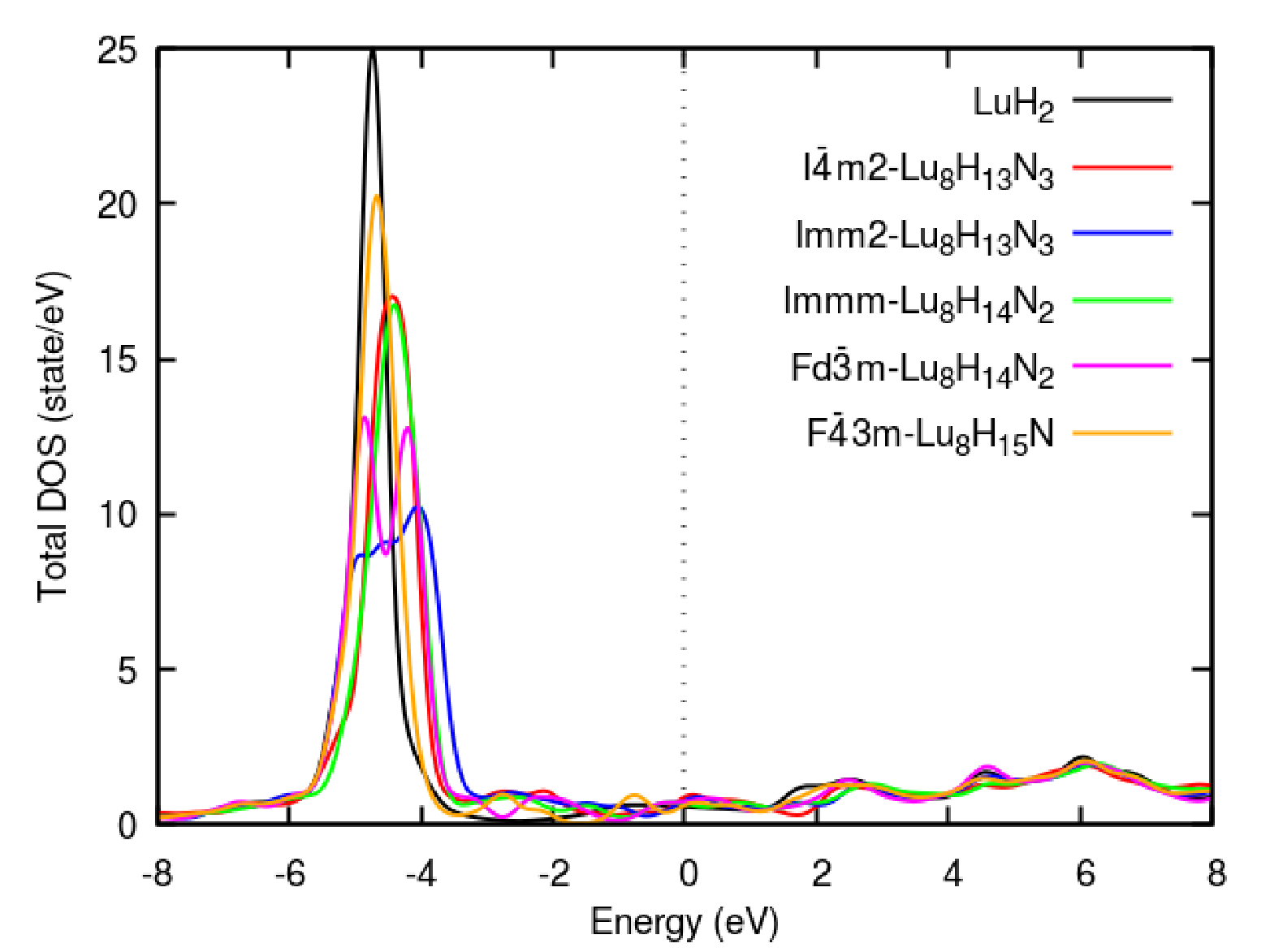}
    \includegraphics[width=0.48\textwidth]{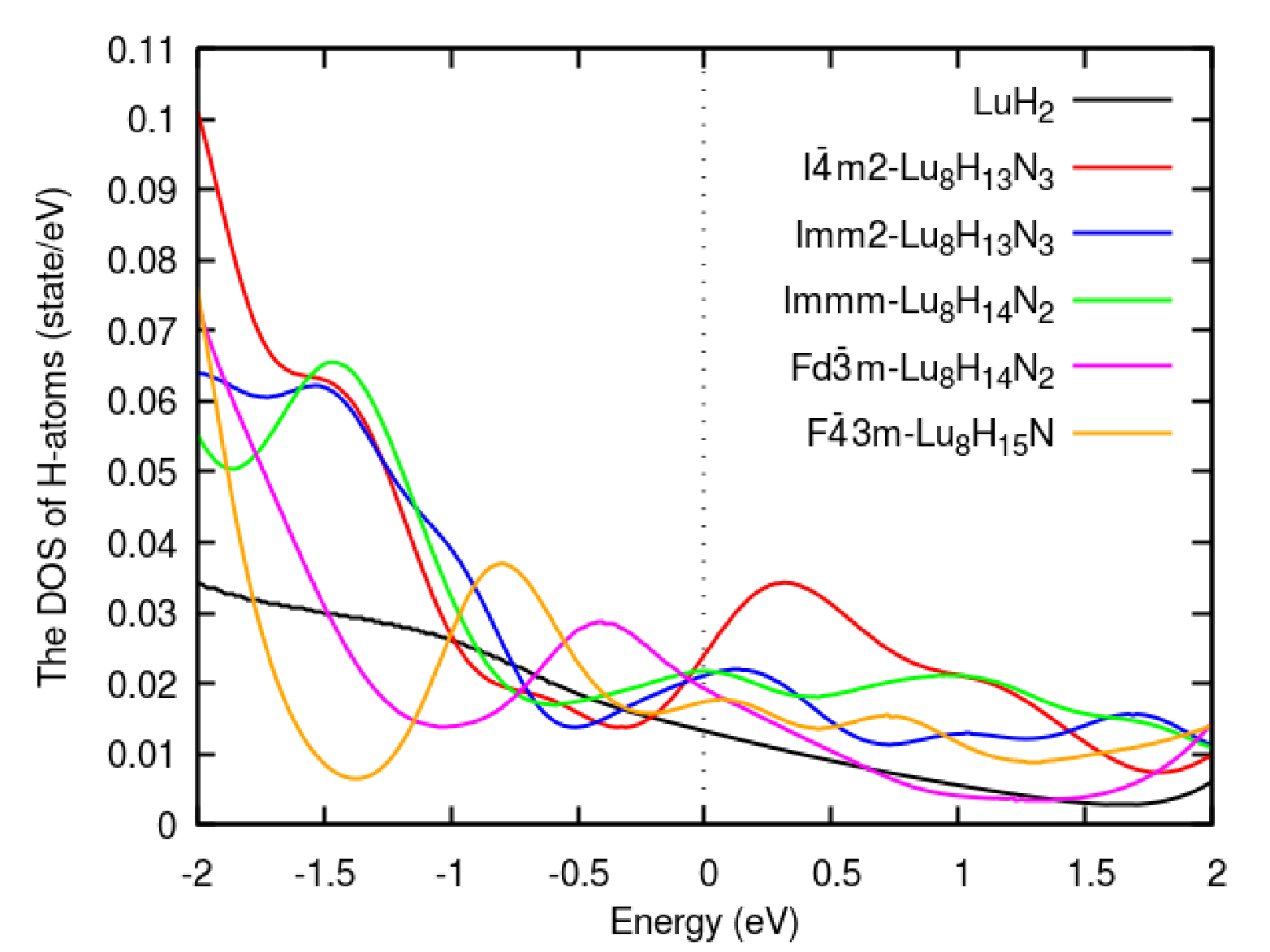}
    \caption{The total DOS and the DOS of H-atoms of LuH$_2$, in comparison with that of Lu$_{8}$H$_{13}$N$_3$, Lu$_{8}$H$_{14}$N$_2$ and Lu$_{8}$H$_{15}$N at 2 GPa.}
    \label{DOS.N-doping}
\end{figure*}